\documentclass[prd,preprint,showpacs,preprintnumbers,nofootinbib,eqsecnum,superscriptaddress]{revtex4}

 \usepackage[dvips,final]{graphicx}
  \usepackage{amssymb}
   \usepackage{amsmath}
    \usepackage{amsfonts}
     \usepackage{epsfig}
      \usepackage{bm}

\usepackage{mathpazo}

\usepackage[section]{placeins}

\usepackage{multirow}
\usepackage{ctable}
\usepackage{booktabs}
\usepackage{array}
\usepackage{tabularx}
\usepackage{xcolor}
\usepackage{pstricks}

\begin{document}

\title{\bm{$D$} meson production asymmetry, unfavoured fragmentation \\
and consequences for prompt atmospheric neutrino production}

\author{Rafa{\l} Maciu{\l}a}
\email{rafal.maciula@ifj.edu.pl} \affiliation{Institute of Nuclear
Physics, Polish Academy of Sciences, Radzikowskiego 152, PL-31-342 Krak{\'o}w, Poland}

\author{Antoni Szczurek\footnote{also at University of Rzesz\'ow, PL-35-959 Rzesz\'ow, Poland}}
\email{antoni.szczurek@ifj.edu.pl} \affiliation{Institute of Nuclear
Physics, Polish Academy of Sciences, Radzikowskiego 152, PL-31-342 Krak{\'o}w, Poland}

\begin{abstract}
We consider unfavoured light quark/antiquark to $D$ meson fragmentation.
We discuss nonperturbative effects for small transverse momenta.
The asymmetry for $D^+$ and $D^-$ production measured by the LHCb 
collaboration provides natural constraints on the parton 
(quark/antiquark) fragmentation functions.
We find that already a fraction of $q/{\bar q} \to D$ fragmentation
probability is sufficient to account for the measured asymmetry. 
We make predictions for similar asymmetry for neutral $D$ mesons.
Large $D$-meson production asymmetries are found for
large $x_F$ which is related to dominance of light quark/antiquark
$q/\bar q \to D$ fragmentation over the standard $c \to D$ fragmentation. As a consequence, prompt atmospheric neutrino 
flux at high neutrino energies can be much larger than for 
the conventional $c \to D$ fragmentation. 
The latter can constitute a sizeable background for the cosmic
neutrinos claimed to be observed recently by the IceCube Observatory. 
Large rapidity-dependent $D^+/D^-$ and $D^0/{\bar D}^0$ asymmetries 
are predicted for low ($\sqrt{s} =$ 20 - 100 GeV) energies. 
The $q/\bar q \to D$ fragmentation leads to enhanced 
production of $D$ mesons at low energies.
At $\sqrt{s}$ = 20 GeV the enhancement factor with respect to
the conventional contribution is larger than a factor of five.
In the considered picture the large-$x_F$ $D$ mesons 
are produced dominantly via fragmentation of light quarks/antiquarks.
Predictions for fixed target $p+^{4}\!\textrm{He}$ collisions relevant 
for a fixed target LHCb experiment are presented.
\end{abstract}

\maketitle

\section{Introduction}

It is believed that the $D$ mesons are produced dominantly
via $c \to D$ fragmentation.
However, asymmetries for $D^+$ and $D^-$ production were obtained
at lower energies for $\pi^-$-nucleus collisions \cite{exp_data_pim} 
and $\Sigma^-$-nucleus collisions \cite{exp_data_Sigmam} and 
recently at the LHC for proton-proton collisions \cite{LHCb_asymmetry}.
Rather small asymmetries of the order of 1\% were found by 
the LHCb collaboration \cite{LHCb_asymmetry}.
One can believe in such low asymmetries as the $CP$ asymmetries
in decay defined as:
\begin{equation}
A_{CP} = \frac{\Gamma(D^+) - \Gamma(D^-)}{\Gamma(D^+) + \Gamma(D^-)}
\label{A_CP}
\end{equation}
were found to be extremely small, consistent with zero (see \textit{e.g.} Refs.~\cite{Babar_A_CP,Belle_A_CP,LHCb_A_CP} and references therein).
The LHCb result was obtained for $D^{\pm} \to K_s^0 K^{\pm}$ decays.

Can perturbative effects lead to any asymmetry?
Higher-order pQCD and electroweak effects on $c \bar c$ asymmetry 
(both quark and antiquark registered) was studied in Ref.~\cite{GHPR2015} 
for $E_T >$ 20 GeV.
The predicted effect was, however, rather small ($<$ 1 \%), at least
for the LHCb (pseudo)rapidity coverage 2 $< \eta <$ 4.

The production asymmetries were interpreted in Refs.~\cite{CGNN2013,HW2016} 
as due to meson cloud mechanism and specific structure of the proton Fock
components. The string model approach to the problem of heavy meson
production and asymmetries in the production of heavy mesons 
was discussed in extent in Ref.~\cite{NS2000}. The LHCb asymmetry was
discussed also in the framework of heavy-quark recombination approach
\cite{LLP2014} (for earlier work see \textit{e.g.} Ref.~\cite{RS2003}).
Here there are four unknown parameters responsible for formation of
$D$ mesons. It was shown that with some combination of parameters
one can describe the LHCb data \cite{LLP2014}.

The conventional $D$ meson production mechanism leads to
symmetry in $D^+/D^-$ or $D^0/{\bar D}^0$ production,
\textit{i.e.} $\sigma(D^+) = \sigma(D^-)$ and 
$\sigma(D^0) = \sigma(\bar{D}^0)$.
As will be discussed in the present paper, only a subtle
isospin-violating effect in vector $D$ meson decays ($D^* \to D X$)
leads to a significant effect of $\sigma(D^+/D^-) < \sigma(D^0/{\bar D}^0)$.

Here we consider a simple alternative phenomenological explanation 
using so-called unfavored fragmentation functions responsible for 
light quark/antiquark fragmentation to $D$ mesons.
Such unfavoured fragmentation functions are known to be important
\textit{e.g.} for $K^+/K^-$ production and corresponding asymmetries
obtained at SPS \cite{SPS} and RHIC \cite{Brahms}. 
Such asymmetries for kaon production were nicely explained in 
the picture of subleading parton fragmentation at low energies \cite{Czech}.
The unfavoured fragmentation functions $g \to D$, $q/{\bar q} \to D$ 
that fullfil DGLAP equations were discussed \textit{e.g.} in 
Ref.~\cite{KKKS2008}.
Even assuming that at the initial scale the fragmentation functions
vanish, they naturally appear at larger scales. The parameters 
of fragmentation functions were
found in some fits to the $e^+ e^-$ data \cite{KKKS2008}.
It is interesting whether the so-obtained unfavoured fragmentation 
functions can describe the observed experimentally asymmetries 
in proton-proton collisions.

In the present paper we wish to constrain the strength of
$q \to D$ ($\bar q \to D$) fragmentation functions using the recent
LHCb data for $D^+$/$D^-$ asymmetry.
Then we shall discuss $q/{\bar q} \to D^{\pm}$ contribution 
to $d \sigma/d x_F$ distributions.
Possible consequences for lower energies and/or for prompt atmospheric 
neutrino production will be discussed.

\section{A theoretical basis of the present calculations}

In this section we briefly review basic ingredients needed
in the present analysis.

\subsection{Light quark/antiquark production}

We start with high collision energies.
We calculate the dominant at large $x_F$ high-energy processes:
$u g \to u g$, $d g \to d g$, $\bar u g \to \bar u g$ and 
$\bar d g \to \bar d g$ and subsequent light quark/antiquark to 
D meson fragmentation and/or decays. The calculations are done in 
the leading-order (LO) collinear factorization approach with a special
treatment of minijets at low transverse momenta,
as adopted in \textsc{Pythia}, 
by multiplying standard cross section by a somewhat arbitrary
suppression factor \cite{Sjostrand:2014zea}
\begin{equation}
F_{sup}(p_T) = \frac{p_T^4}{((p_{T}^{0})^{2} + p_T^2)^2} \theta(p_T -
p_{T,cut}) \; .
\label{suppression_factor}
\end{equation}
%

\begin{figure}[!h]
\begin{minipage}{0.47\textwidth}
 \centerline{\includegraphics[width=1.0\textwidth]{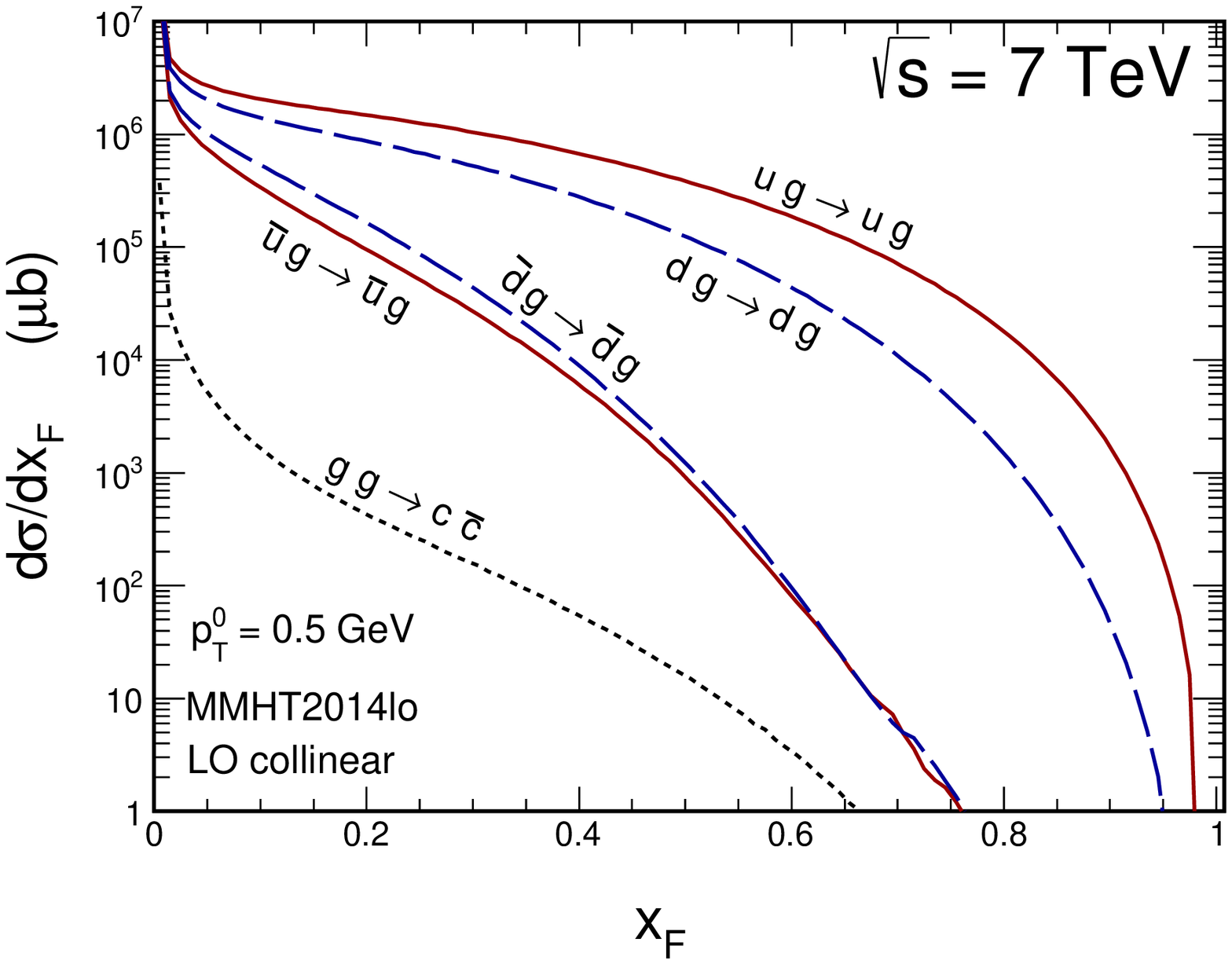}}
\end{minipage}
\hspace{0.5cm}
\begin{minipage}{0.47\textwidth}
 \centerline{\includegraphics[width=1.0\textwidth]{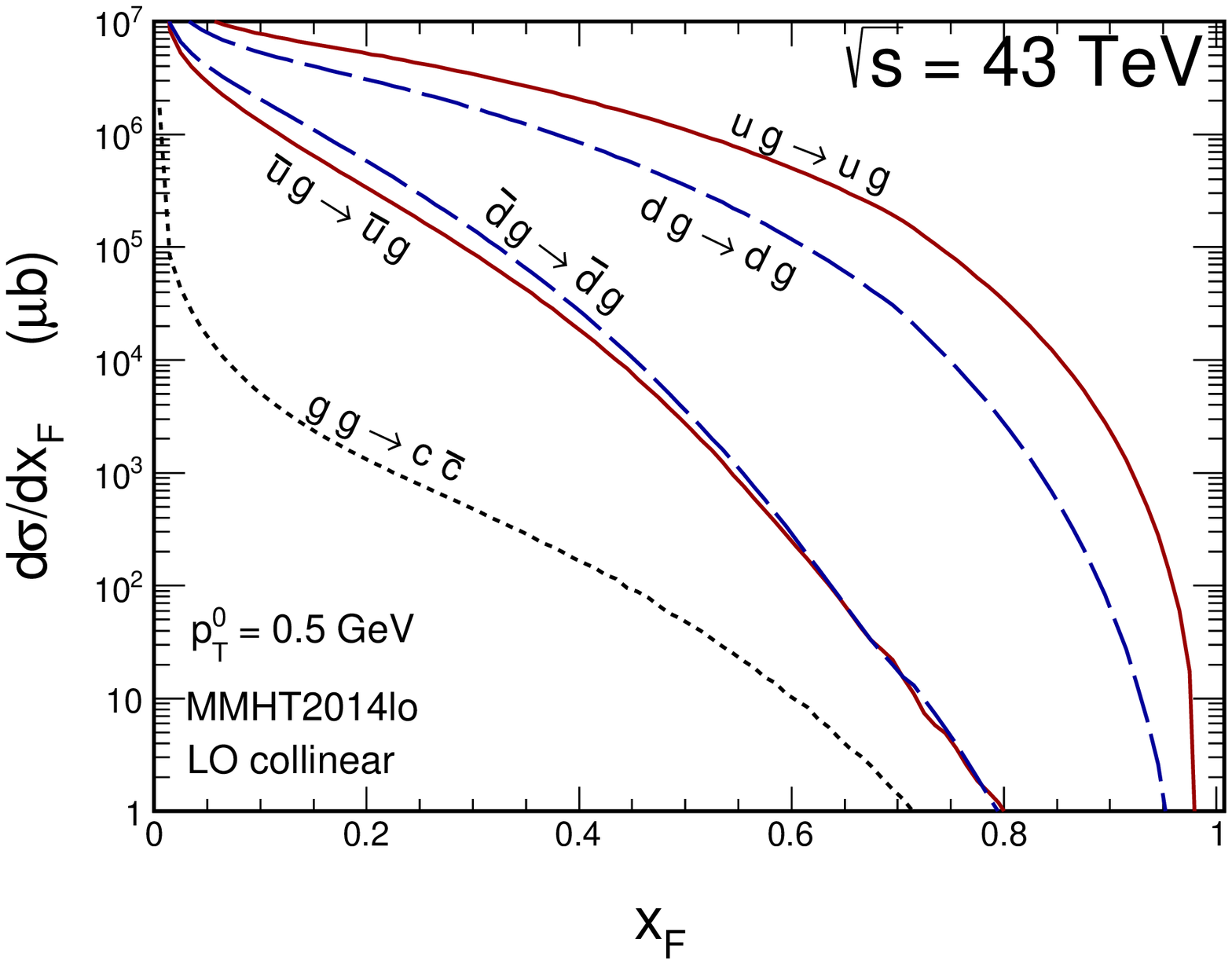}}
\end{minipage}
   \caption{
\small Quark and antiquark distributions in Feynman $x_F$ for 
$\sqrt{s} =$ 7 TeV (left panel) and $\sqrt{s} =$ 43 TeV (right panel)
corresponding to $E_{\mathrm{lab}}(p)$ = 10$^{9}$ GeV (relevant for high-energy
prompt atmospheric neutrinos).
This calculation was performed within collinear-factorization approach
with $p_{T}^{0} =$ 0.5 GeV.
 }
 \label{fig:dsig_dxf_partons}
\end{figure}

\begin{figure}[!h]
\begin{minipage}{0.47\textwidth}
 \centerline{\includegraphics[width=1.0\textwidth]{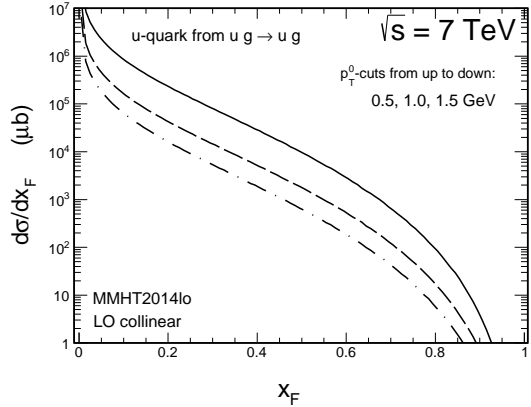}}
\end{minipage}
   \caption{
\small Light u-quark distribution in Feynman $x_F$ for 
$\sqrt{s} =$ 7 TeV for different values of $p_{T}^{0}$ = 0.5, 1.0, and 1.5 GeV.
 }
 \label{fig:dsig_dxf_partons_ptcut}
\end{figure}

First we calculate distributions of $u$, $d$, $\bar u$, $\bar d$
in Feynman $x_F$ in the forward (projectile) region.
In Fig.~\ref{fig:dsig_dxf_partons} we show distributions in $x_F$ of 
the light-quarks/antiquarks obtained in the collinear-factorization 
approach.
In this calculation we use the MMHT2014lo \cite{Harland-Lang:2014zoa} parton distributions.
The factorization and renormalization scales are taken as:
$\mu_F^2, \mu_R^2 = \mu_0^2 + p_T^2$.
Here we take $\mu_0^2$ = 0.5$^2$ GeV$^2$.
In Fig.~\ref{fig:dsig_dxf_partons_ptcut} we show results for different
values of $p_{T}^{0}$ = 0.5, 1.0, and 1.5 GeV.
We think that already with $p_{T}^{0}$ = 0.5 GeV reliable quark/antiquark 
distributions in $y$ and $x_F$ are obtained. The shapes for different $p_{T}^{0}$ are rather similar.

\begin{figure}[!h]
\begin{minipage}{0.47\textwidth}
 \centerline{\includegraphics[width=1.0\textwidth]{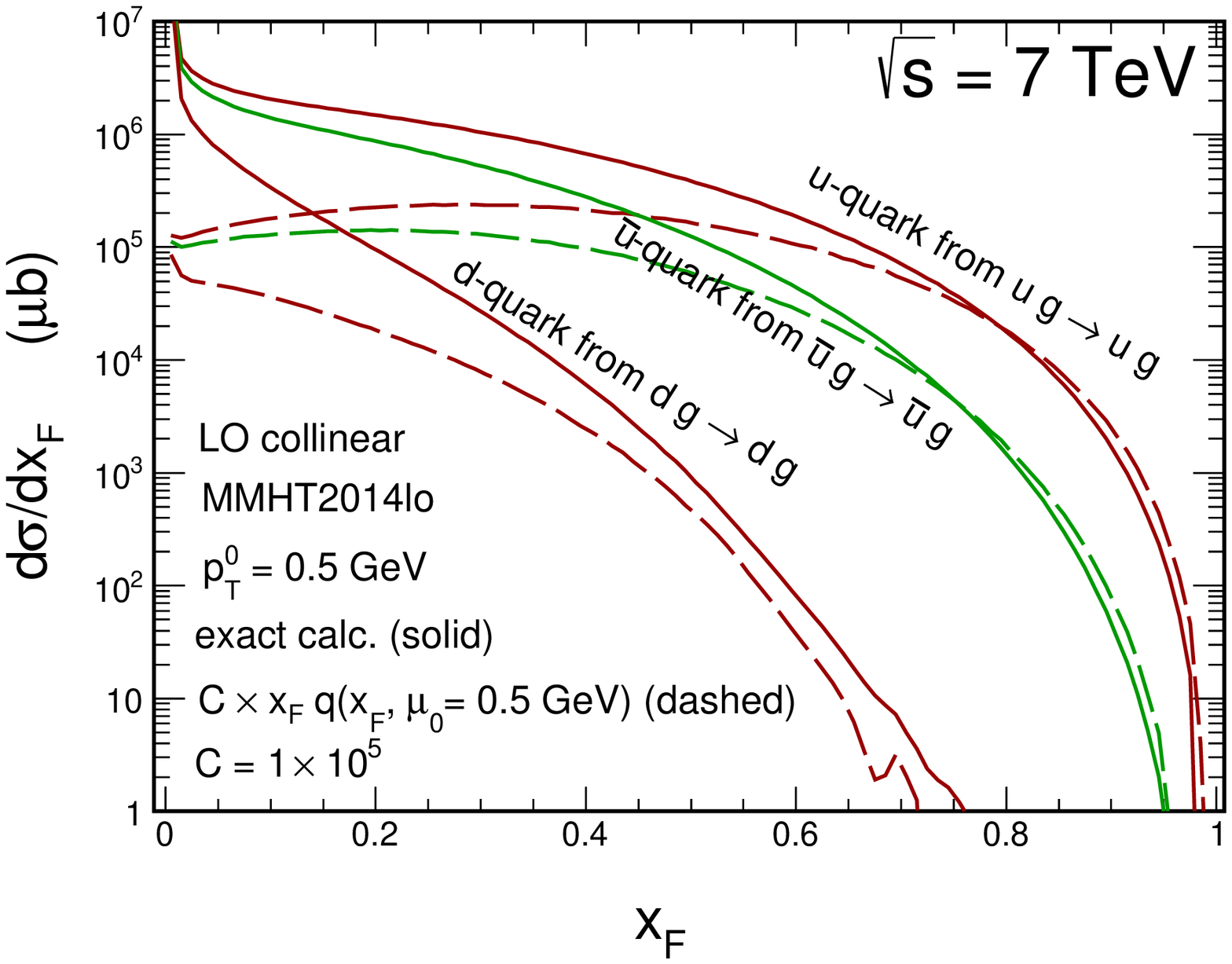}}
\end{minipage}
\begin{minipage}{0.47\textwidth}
 \centerline{\includegraphics[width=1.0\textwidth]{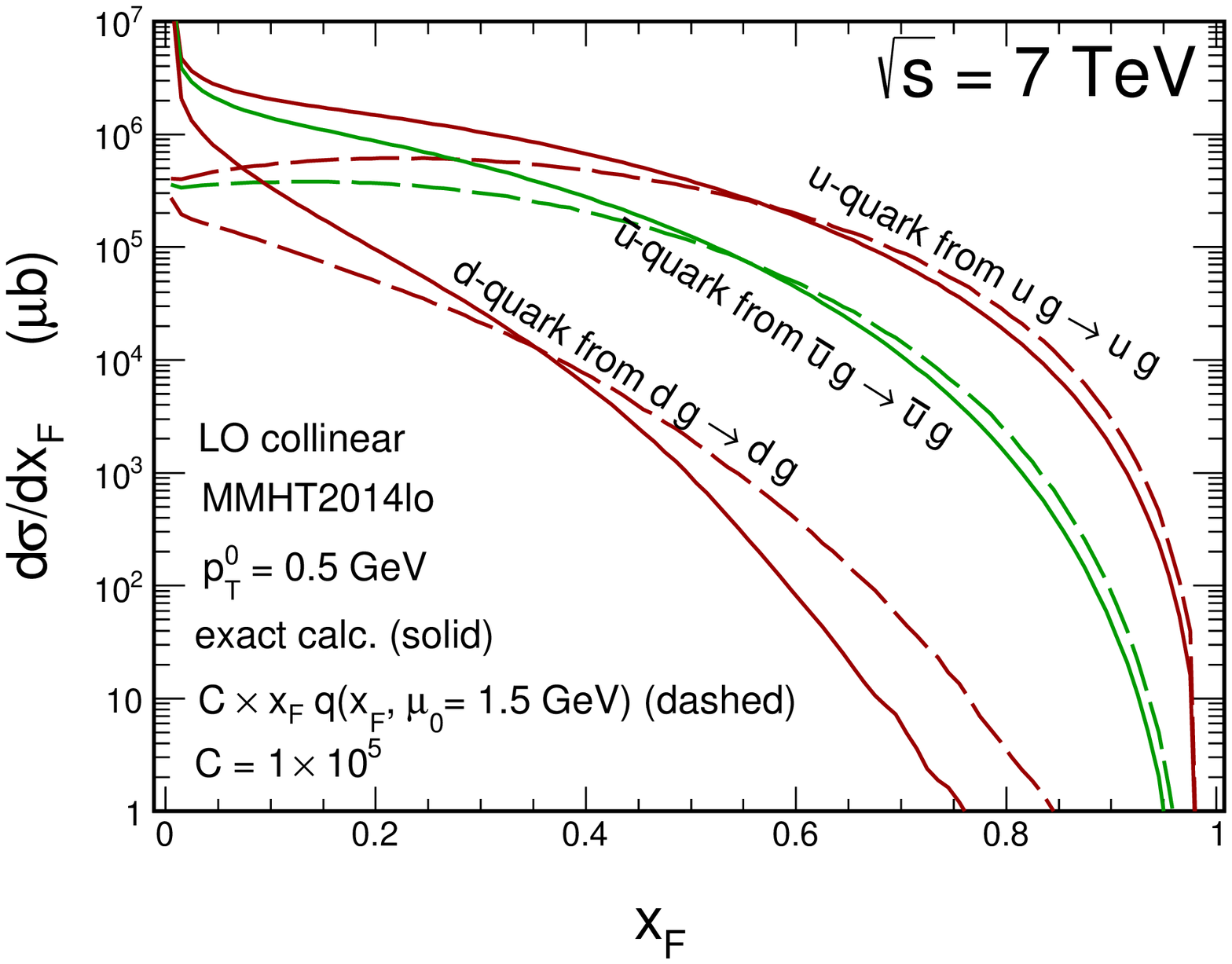}}
\end{minipage}
\begin{minipage}{0.47\textwidth}
 \centerline{\includegraphics[width=1.0\textwidth]{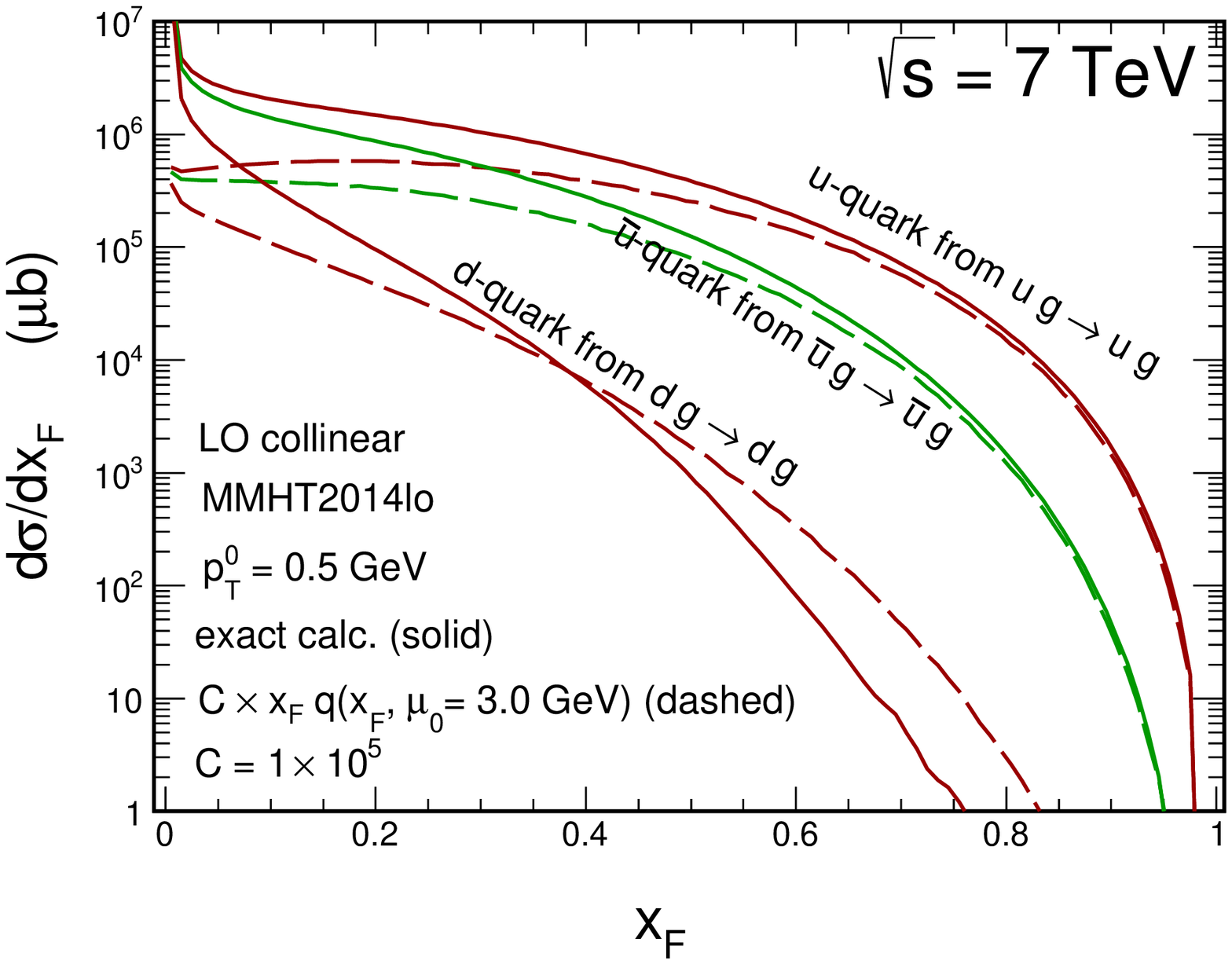}}
\end{minipage}
    \caption{
\small Distribution in Feynman $x_F$ for $u$ and $d$ quarks
and $\bar u$ antiquarks calculated 
with formula (\ref{naive_quark_distributions})
for different factorization scales given explicitly in the figures.
}
\label{fig:dsig_dxf_naive}
\end{figure}

\begin{figure}[!h]
\begin{minipage}{0.47\textwidth}
  \centerline{\includegraphics[width=1.0\textwidth]{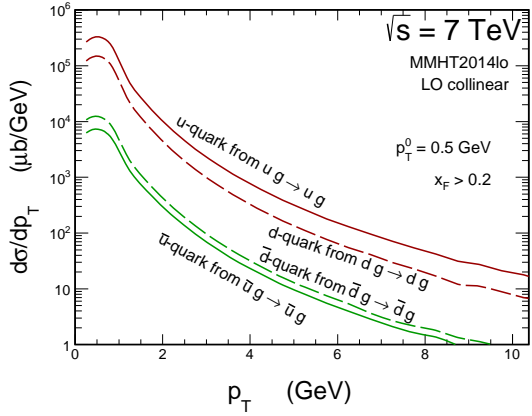}}
\end{minipage}
   \caption{
\small Transverse momentum distribution of light quarks/antiquarks for
       $x_F >$ 0.2.
}
 \label{fig:dsig_dpt_partons}
\end{figure}

At large $x_F$ the distribution of produced quarks/antiquarks
can be approximated in terms of partons in the initial hadron as
\begin{equation}
\frac{d \sigma}{d x_F} \left( x_F \right)
\approx C(\sqrt{s}) x_F q_f (x_F,\mu_{eff}^2) \; ,
\label{naive_quark_distributions}
\end{equation}
where $\mu_{\mathrm{eff}}^2$ is the scale relevant for low transverse momentum 
quark/antiquark production.
In Fig.~\ref{fig:dsig_dxf_naive} we compare results of calculations
performed in the collinear-factorization approach with those obtained with 
the very simple approximation given by Eq.~(\ref{naive_quark_distributions}).
We see a reasonably good agreement of the results of the two calculations. 
The same parton distribution set was used in both cases.
In this calculation $\mu_{\mathrm{eff}} =$ 0.5-3 GeV was taken.
The agreement for $u$ and $d$ quarks is much better than that for 
$\bar u$ and $\bar d$ antiquarks.
The best agreement is obtained for $\mu_{\mathrm{eff}} \approx$ 2-3 GeV.

The dependence on transverse momentum of quarks/antiquarks is very steep.
In Fig.~\ref{fig:dsig_dpt_partons} we show the transverse momentum
distribution of produced light quarks and antiquarks.
Here we have assumed a lower cut on $x_F >$ 0.2 to concentrate
on the interesting for us region related to fast prompt 
atmospheric neutrinos \cite{GMPS2017}.
Althought there is a strong dependence of the cross section on $p_T$ 
the integrated cross sections are finite. The averaged transverse 
momentum is $p_T \sim$ 2 GeV.

\subsection{Unfavoured fragmentation functions}

Let us start with direct fragmentation.
Then we have to include $u,\bar u, d, \bar d \to D^i$ parton
fragmentation. 
The corresponding fragmentation functions fulfill the following
flavour symmetry conditions:
\begin{equation}
D_{d \to D^-}(z) = D_{\bar d \to D^+}(z) = D^{(0)}(z) \; .
\label{ff_symmetries}
\end{equation}
Similar symmetry relations hold for fragmentation 
of $u$ and $\bar u$ to $D^0$ and $\bar D^0$ mesons.
However $D_{q \to D^0}(z) \ne D_{q \to D^+}(z)$ which is caused
by the contributions from decays of vector $D^*$ mesons.
Furthermore we assume for doubly suppressed fragmentations:
\begin{equation}
D_{\bar u \to D^{\pm}}(z) = D_{u \to D^{\pm}}(z) = 0 \; .
\label{neglected_ff}
\end{equation}

The fragmentation functions at sufficiently large scales undergo 
DGLAP evolution equations \cite{KKKS2008}
\begin{equation}
\frac{d}{d \mathrm{ln} \mu^2} D_a(x,\mu) =
\frac{\alpha_s(\mu)}{2 \pi} \sum_b 
\int_x^1 \frac{dy}{y} P_{a \to b}^T \left( y,\alpha_s(\mu) \right) 
D_b\left( \frac{x}{y},\mu \right) \; ,
\label{DGLAP_for_FF}
\end{equation}
where $a = g, u, \bar u, d, \bar d, s, \bar s, c, \bar c$.
In the case of $e^+ e^-$ collisions the scale is usually taken as $\mu^2 = s$.
When fitting fragmentation functions to $e^+ e^- \to D$ data one
usually assumes
\begin{equation}
D_{q/\bar q \to D}(z,\mu_0^2) = 0 
\label{initial condition}
\end{equation}
at some initial scale usually taken as $\mu_0 = m_c, 2 m_c$, where $m_{c}$ is charm quark mass.
This simplification is not a good approximation for the case
of proton-proton collisions where the asymmetry was observed 
\cite{LHCb_asymmetry} even at very low transverse momenta.
Here we are particularly interested in low transverse momentum $D$ mesons.
Then our typical factorization scales $\mu^2 = p_T^2 + m_q^2$ are very
small.
Therefore we limit in the following to a phenomenological approach
and ignore possible DGLAP evolution effects important at somewhat larger
transverse momenta.
We can parametrize the unfavoured fragmentation functions in this
phase space region as:
\begin{equation}
D_{q \to D}(z) = A_{\alpha} (1-z)^{\alpha} \; .
\label{ff_simple_parametrization}
\end{equation}
Instead of fixing the uknown $A_{\alpha}$ we will operate rather with
the fragmentation probability:
\begin{equation}
P_{q \to D} = \int dz \; A_{\alpha} \left( 1 - z \right)^{\alpha} \; .
\label{Dff_simple_parametrization}
\end{equation}
and calculate corresponding $A_{\alpha}$ for a fixed $P_{q \to D}$ and
$\alpha$.
Therefore in our effective approach we have only two free parameters.

Another simple option one could consider is:
\begin{equation}
D_{q_f \to D}(z) = P_{q_f \to D} \cdot D_{\mathrm{Peterson}}(1-z) \; .
\label{Peterson}
\end{equation}
Then $P_{q_f \to D}$ would be the only free parameter.

In addition to the direct fragmentation (given by $D^{(0)}(z)$) 
there are also contributions with intermediate vector $D^*$ mesons.
Then the chain of production of charged $D$ mesons is naively as follows:
\begin{eqnarray}
&&\bar u \to D^{*,0} \to D^+ \; \mathrm{(forbidden)}, \nonumber \\
&&     u \to {\bar D}^{*,0} \to D^- \; \mathrm{(forbidden)}, \nonumber \\
&&\bar d \to D^{*,+} \to D^+ \; \mathrm{(allowed)}, \nonumber \\
&&     d \to D^{*,-} \to D^- \; \mathrm{(allowed)}.
\label{intermediate_vectors}  
\end{eqnarray}
In reality the first two chains are not possible as the decays 
of corresponding vector mesons ($D^{*,0}$ and $\bar D^{*,0}$) 
are forbidden by lack of phase space.
This would be, however, possible for $D^0$ and $\bar D^0$ production
where $D^{*,\pm}$ may decay producing $D^0$ or $\bar D^0$ mesons.
In the latter case the two terms have different flavour structure
and the production asymmetry is more complicated.
In addition $D^0$-$\bar D^0$ oscillations occur 
(see \textit{e.g.} Refs.~\cite{LHCb_oscillations1,LHCb_oscillations2}) which makes 
the extraction of initial $D^0/{\bar D}^0$ production asymmetry 
a bit more difficult.
According to our knowledge this was not studied so far by the LHCb
collaboration.

Including both direct and resonant contributions the combined 
fragmentation function of light quarks/antiquarks to charged $D$ mesons 
can be written as:
\begin{equation}
D_{d/\bar d \to D^{\mp}}^{\mathrm{eff}}(z) =
D_{d/\bar d \to D^{\mp}}^0(z) +
P_{\mp \to \mp} \cdot D_{d/\bar d \to D^{*,\mp}}^1(z) \; .
\label{charged_D_mesons}
\end{equation}
The decay branching ratios can be found in Ref.~\cite{PDG}
and is $P_{\pm \to \pm} =$ 0.323.
The indirect vector meson contributions have the same flavour structure
as the direct one.
It is easy to check that the decay $D^* \to DX$ practically does not
change the distribution in $z$.

For neutral $D$ mesons we have similarly:
\begin{eqnarray}
D_{u/{\bar u} \to {\bar D}^0/D^0}^{\mathrm{eff}}(z) &=&
D_{u/{\bar u} \to {\bar D}^0/D^0}^0(z) 
+ P_{0 \to 0}  \cdot D_{u/{\bar u} \to {\bar D}^{*,0}/D^{*,0}}^1(z) \; , \\
D_{d/{\bar d} \to {\bar D}^0/D^0}^{\mathrm{eff}}(z) &=& 
  P_{\pm \to 0} \cdot D_{d/{\bar d} \to D^{*,\mp}}^1(z) \; .
\label{neutral_D_mesons}
\end{eqnarray}
Here there are more possibilities than for charged $D$ mesons
as both charged and neutral vector mesons decay into neutral $D$ mesons.
The decay probablities that appeared above are:
$P_{0 \to 0}$ = 0.667 and $P_{\pm \to 0}$ = 1 \cite{PDG}.

We assume flavour symmetry of fragmentation functions also 
for vector $D$ meson production:
\begin{equation}
D_{u/{\bar u} \to {\bar D}^{*,0}/D^{*,0}}^1(z)
= D_{d/{\bar d} \to D^{*,\mp}}^1(z) = D^{(1)}(z) \; .
\label{flavour_symmerty_for_vector}
\end{equation}
Finally we shall take an approximation:
\begin{equation}
D^{(0)}(z) \approx D^{(1)}(z) 
\end{equation}
which can be easily relaxed if needed.
We think that such an approximation is, however, sufficient for 
the present exploratory calculations.

\subsection{\bm{$D$} meson distributions}

At forward directions (relevant for LHCb or IceCube) the details of 
hadronization are fairly important.
Here the hadronization is done as in Ref.~\cite{Czech} assuming
that the hadron pseudorapidity is equal to parton pseudorapidity and
only momenta of hadrons are reduced compared to the parent partons.

In such an approximation the $D$ meson $x_F$-distributions at large 
$x_F$ can be obtained from the quark/antiquark distributions calculated 
in the collinear or $k_t$-factorization approaches as:
\begin{equation}
\frac{d\sigma}{d x_F} = \sum_f
\int_{x_F}^1 \frac{dz}{z}   
\frac{d \sigma(x_F/z)}{d x'_F} D_{q_f \to D}(z) \; .
\label{convolution}
\end{equation}

Instead of the more complicated calculations within collinear or 
$k_t$-factorization one can make first a simplified calculation.
At very small transverse momenta and forward directions ($x_F >$ 0.2) 
the outgoing quarks/antiquarks practically carry the same momentum 
fractions as the initial ones in the proton. 
Approximately one can therefore write the $x_F$- distribution of 
outgoing quarks/antiquarks as
\begin{equation}
\frac{d \sigma_{p p \to D X}(x_F)}{d x_F} \approx C \sum_f
\int_{0}^{1} dz 
\left(x_f/z\right) q_f\left(x_F/z,\mu^2 \right)
\; D_{q_f \to D}(z) \; .
\end{equation}
The constant $C$ is responsible for the cross section normalization
and depends on collision energy $C = C(\sqrt{s})$.
The constant can be fitted to the asymmetries in
experiments that measured different species of $D$ mesons.

\subsection{Flavour asymmetry}

The flavour asymmetry in production is defined as:
\begin{equation}
A_{D^+/D^-}(\xi) 
= \frac{ \frac{d \sigma_{D^-}}{d \xi}(\xi) - \frac{d \sigma_{D^+}}{d \xi}(\xi) }
       { \frac{d \sigma_{D^-}}{d \xi}(\xi) + \frac{d \sigma_{D^-}}{d \xi}(\xi) }
\; ,
\label{asymmetry}
\end{equation}
where $\xi = x_F, y, p_T, (y,p_T)$.
In the following we shall consider several examples of selecting $\xi$.

To calculate asymmetry we have to include also dominant contribution
corresponding to conventional $c/{\bar c} \to D/\bar{D}$ fragmentation.
The leading-order pQCD calculation is not reliable in this context. 
In the following the conventional contribution is calculated within 
the $k_t$-factorization approach with the Kimber-Martin-Ryskin unintegrated parton distributions
\cite{MS_D_mesons} which has proven to well describe the LHC data.
Such an approach seems consistent with collinear next-to-leading order
approach (see \textit{e.g.} a discussion in Ref.~\cite{MS2016}).

\begin{figure}[!h]
\begin{minipage}{0.47\textwidth}
  \centerline{\includegraphics[width=1.0\textwidth]{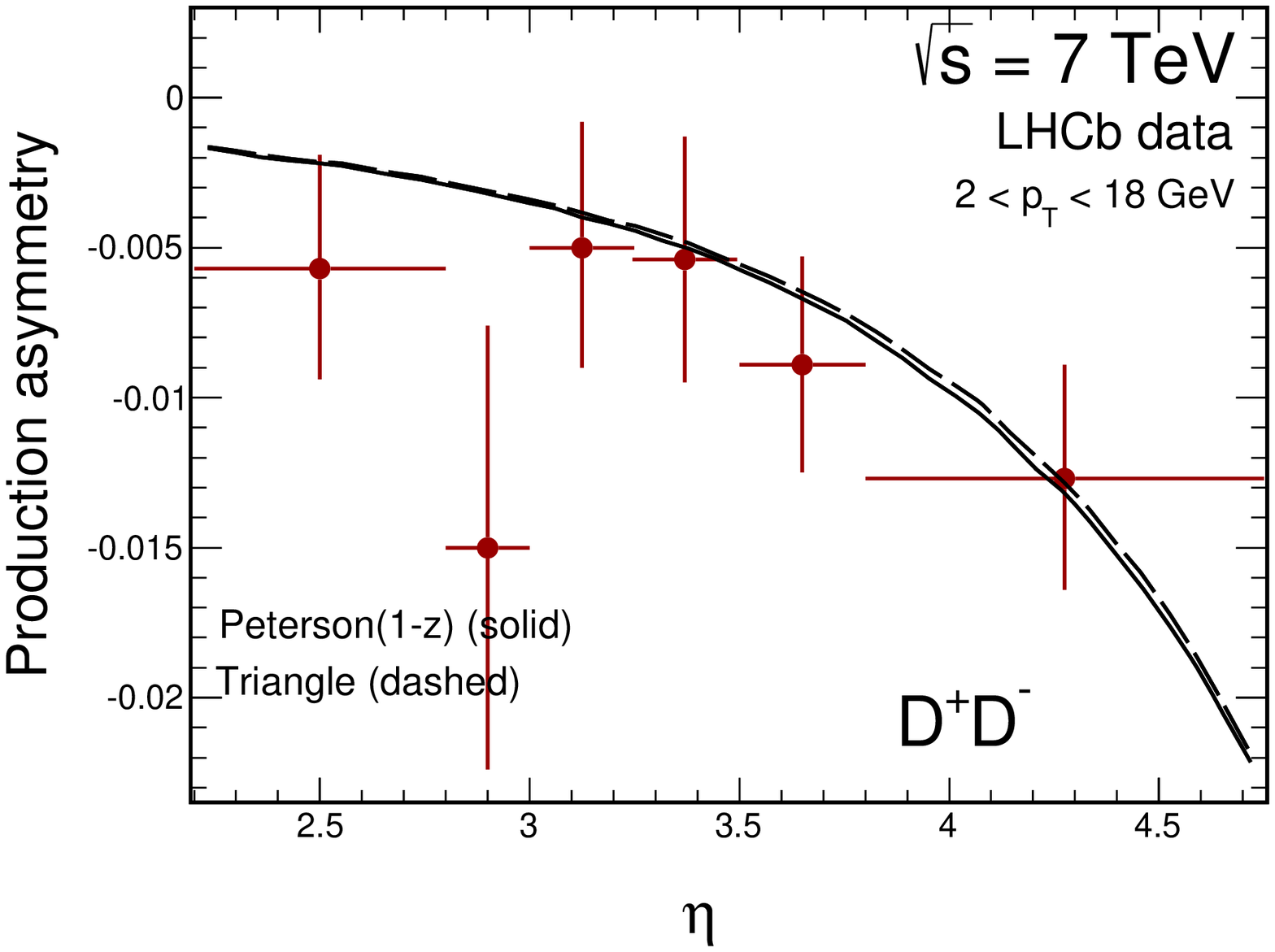}}
\end{minipage}
\hspace{0.5cm}
\begin{minipage}{0.47\textwidth}
  \centerline{\includegraphics[width=1.0\textwidth]{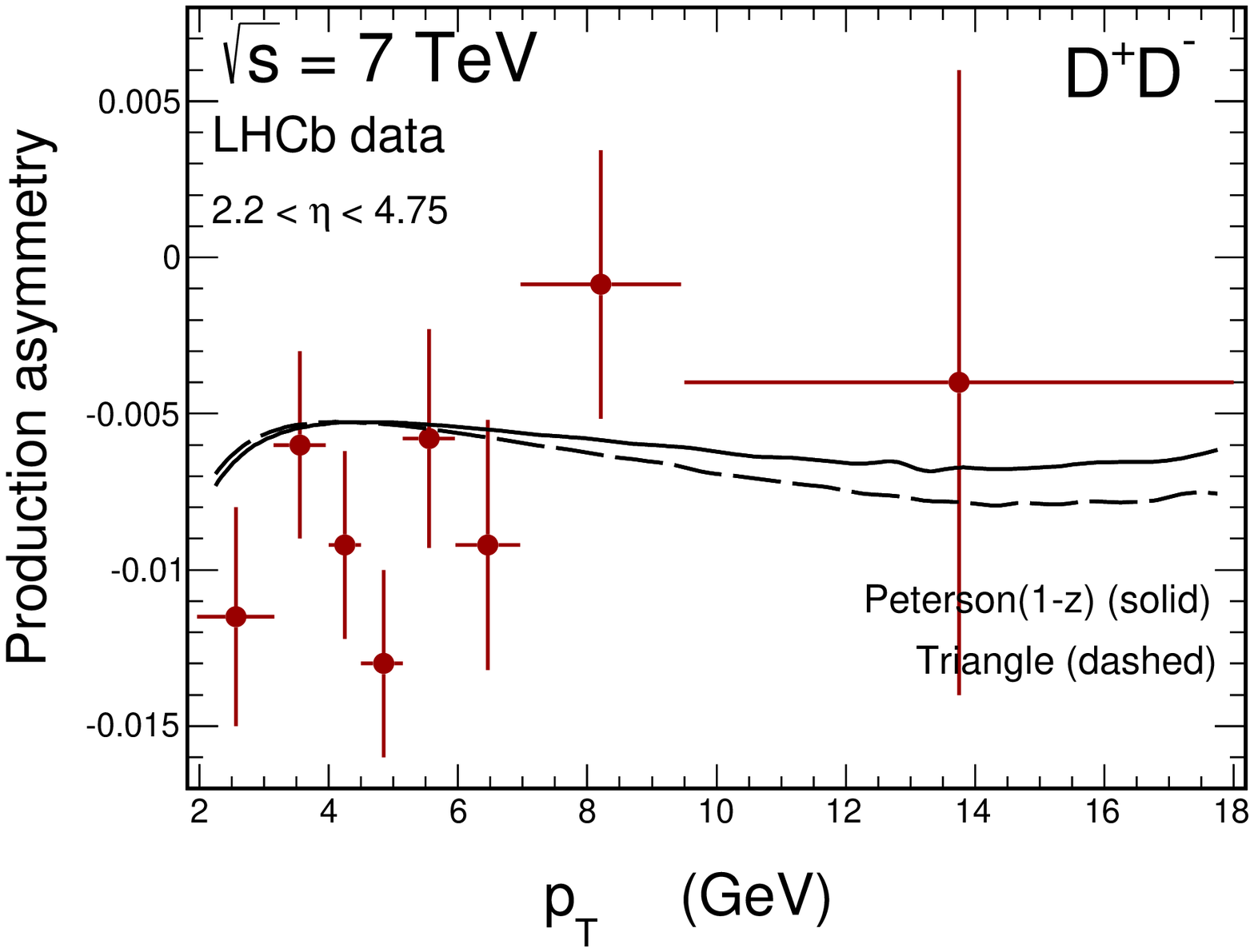}}
\end{minipage}
\begin{minipage}{0.47\textwidth}
  \centerline{\includegraphics[width=1.0\textwidth]{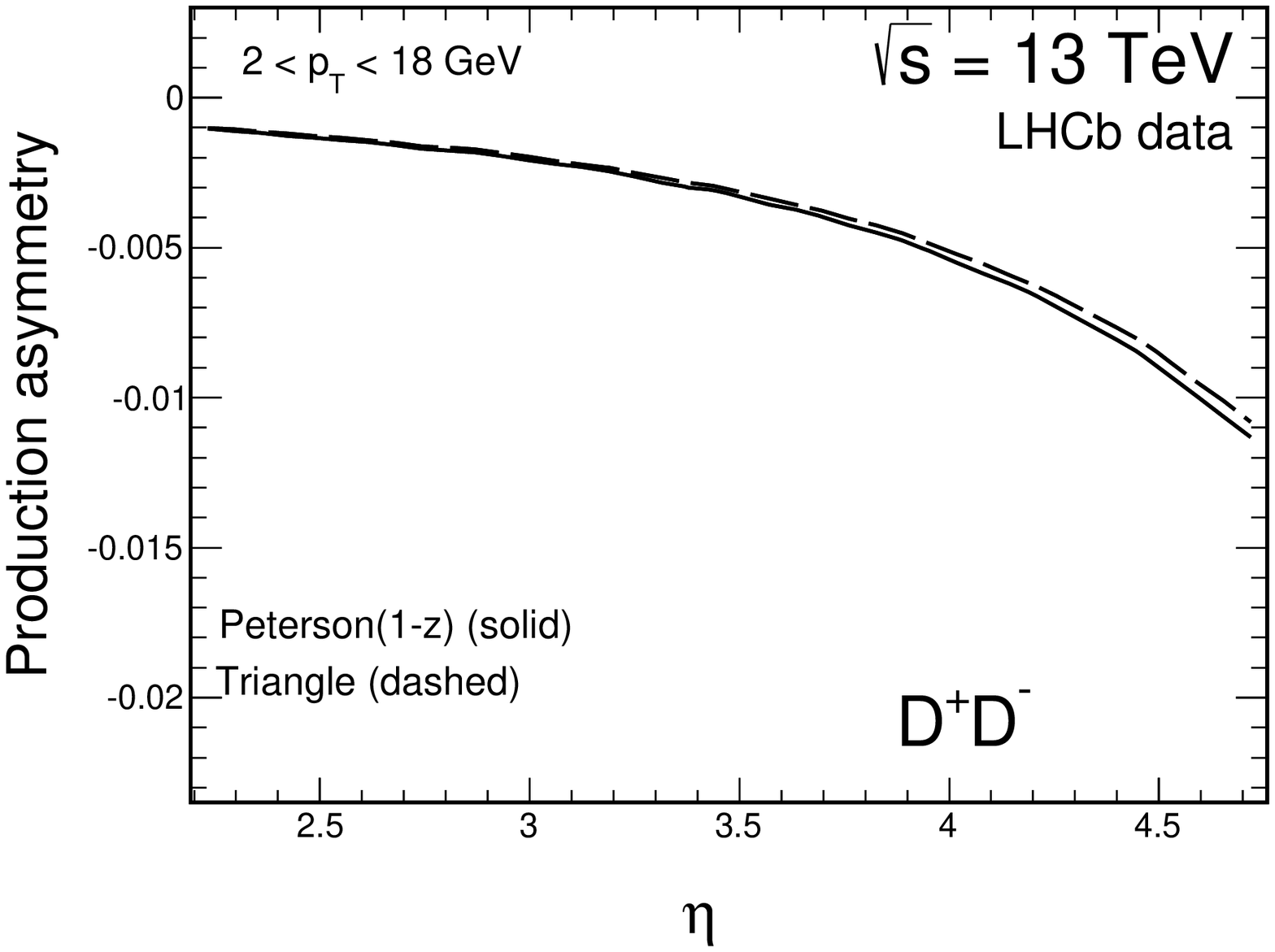}}
\end{minipage}
\hspace{0.5cm}
\begin{minipage}{0.47\textwidth}
  \centerline{\includegraphics[width=1.0\textwidth]{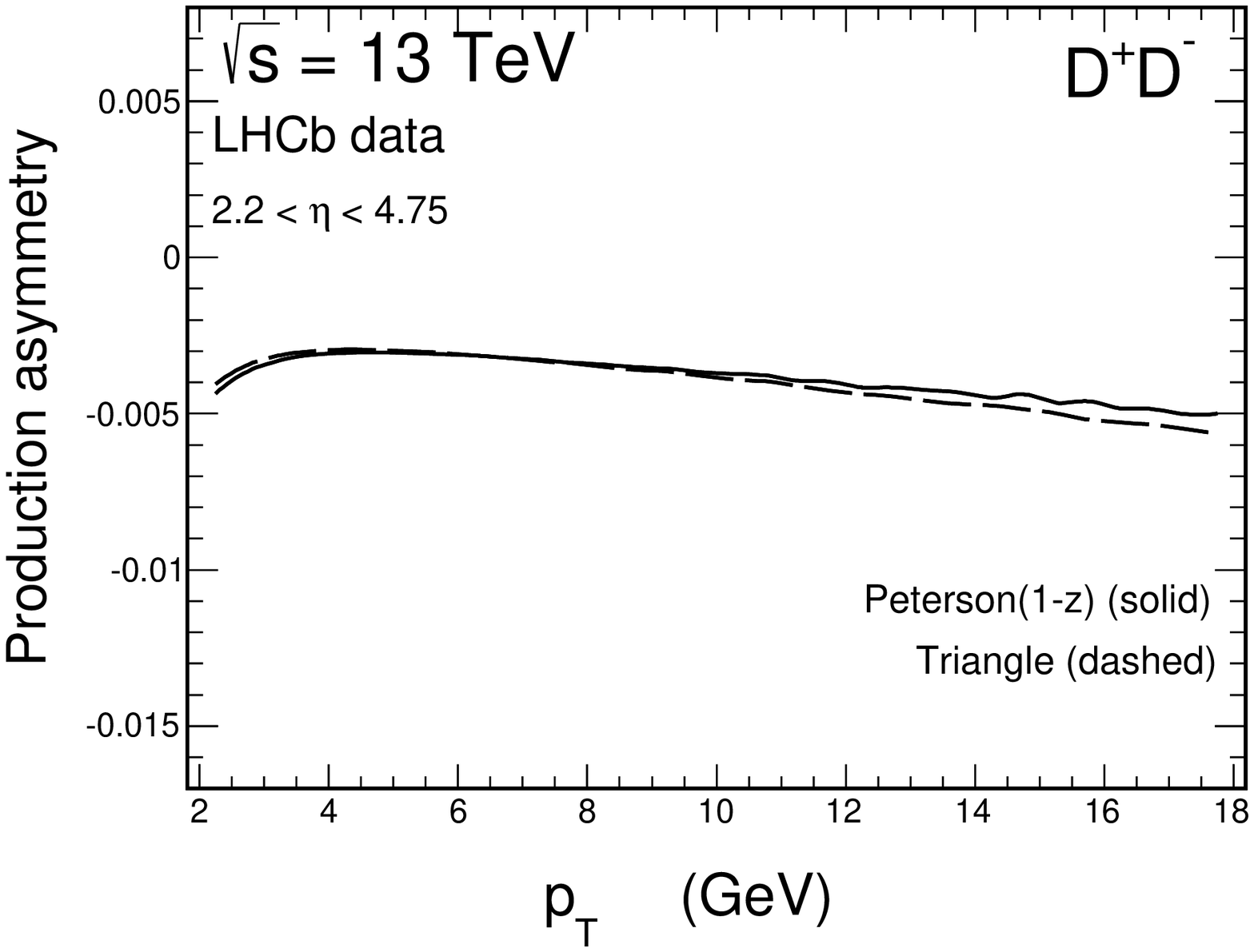}}
\end{minipage}
  \caption{
\small $A_{D^+/D^-}$ production asymmetry measured by the LHCb collaboration at $\sqrt{s}= 7$ TeV as
a function of $D$ meson pseudorapidity (left-top panel) and $D$ meson
transverse momentum (right-top panel). The corresponding predictions for
$\sqrt{s}= 13$ TeV are shown 
in the bottom panels.
}
\label{fig:LHCb_asymmetry_charged}
\end{figure}

For example in top panels of Fig.~\ref{fig:LHCb_asymmetry_charged} we show results for 
the asymmetry for $P_{q \to D}$ adjusted to the LHCb
data. In this calculation, and in the rest of the paper, we have 
fixed $\alpha$ = 1 in formula (\ref{ff_simple_parametrization}).
We shall call corresponding fragmentation functions as triangular for brevity.
In the left panel we show $A_{D^+/D^-}(\eta)$ for $p_{T,D} \in$ (2,18) GeV
and in the right panel we show $A_{D^+/D^-}(p_T)$ for 2.2 $< \eta <$ 4.75 .
We find that $P_{q \to D} =$ 0.005 $\pm$ 0.001 for triangle fragmentation function and $P_{q \to D} =$ 0.007 $\pm$ 0.001 for Peterson(1-z) is consistent with 
main trends of the LHCb data. This are rather small numbers compared
to $c/{\bar c} \to D/{\bar D}$ fragmentation which happens with probability
of the order of 50 \%. The results do not depend on transverse momentum
cut $p_{T}^0$, since the LHCb kinematics excludes the uncertain region 
of very small meson transverse momenta. In the bottom panels we show our predictions for $\sqrt{s}=13$ TeV.

Charm conservation in strong processes must unavoidably lead to extra
$c$ or $\bar c$ production at lower $x_F$ emitted
rather in the remnant direction. The extra emissions lead to a reduction of
asymmetries and enhanced production of charm (both mesons and baryons).
This effect is not included explicitly when fitting the LHCb asymmetries. 
In our opinion the fit includes, however, this effect 
in an effective way.

\begin{figure}[!h]
\begin{minipage}{0.47\textwidth}
  \centerline{\includegraphics[width=1.0\textwidth]{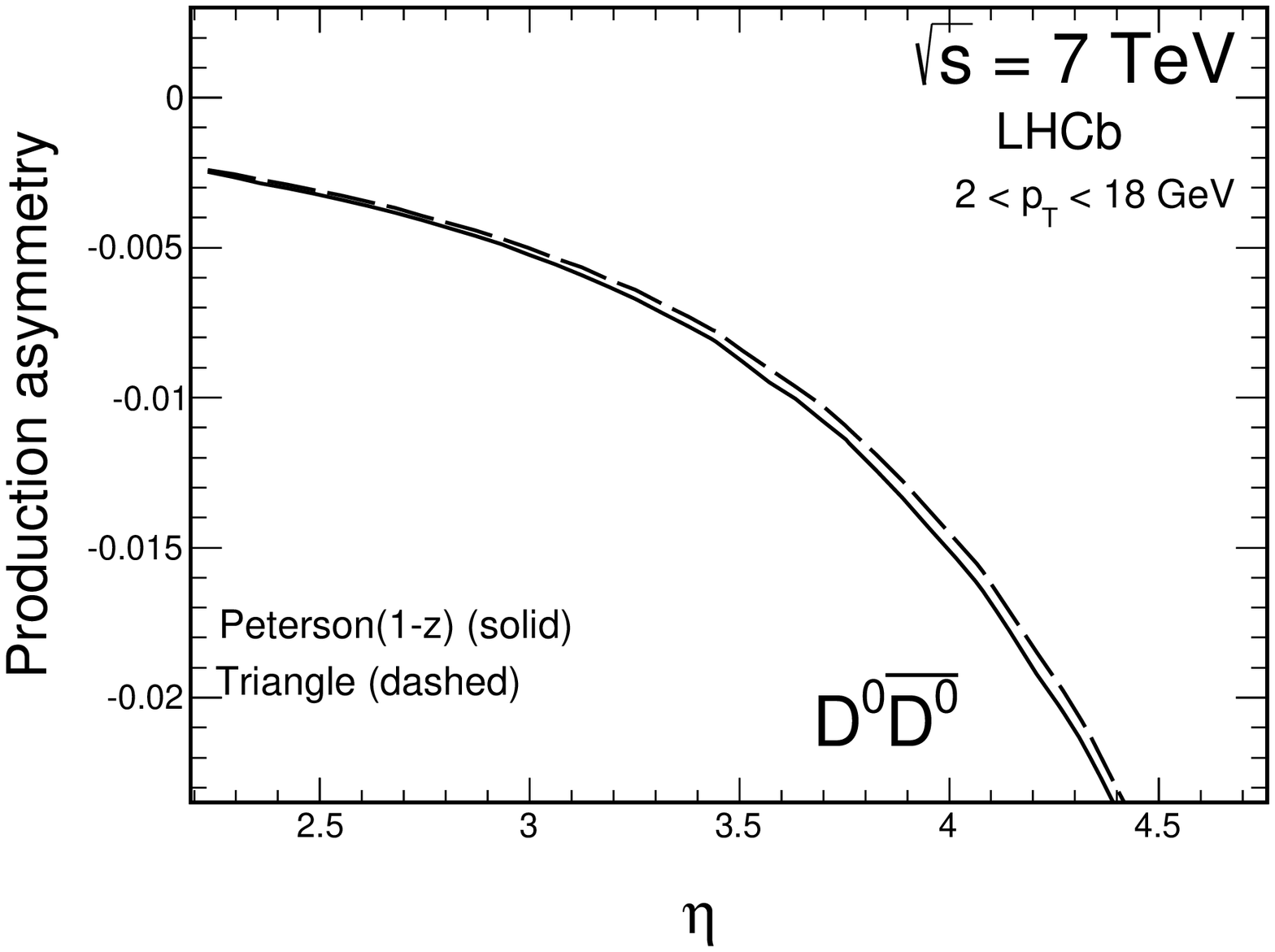}}
\end{minipage}
\hspace{0.5cm}
\begin{minipage}{0.47\textwidth}
  \centerline{\includegraphics[width=1.0\textwidth]{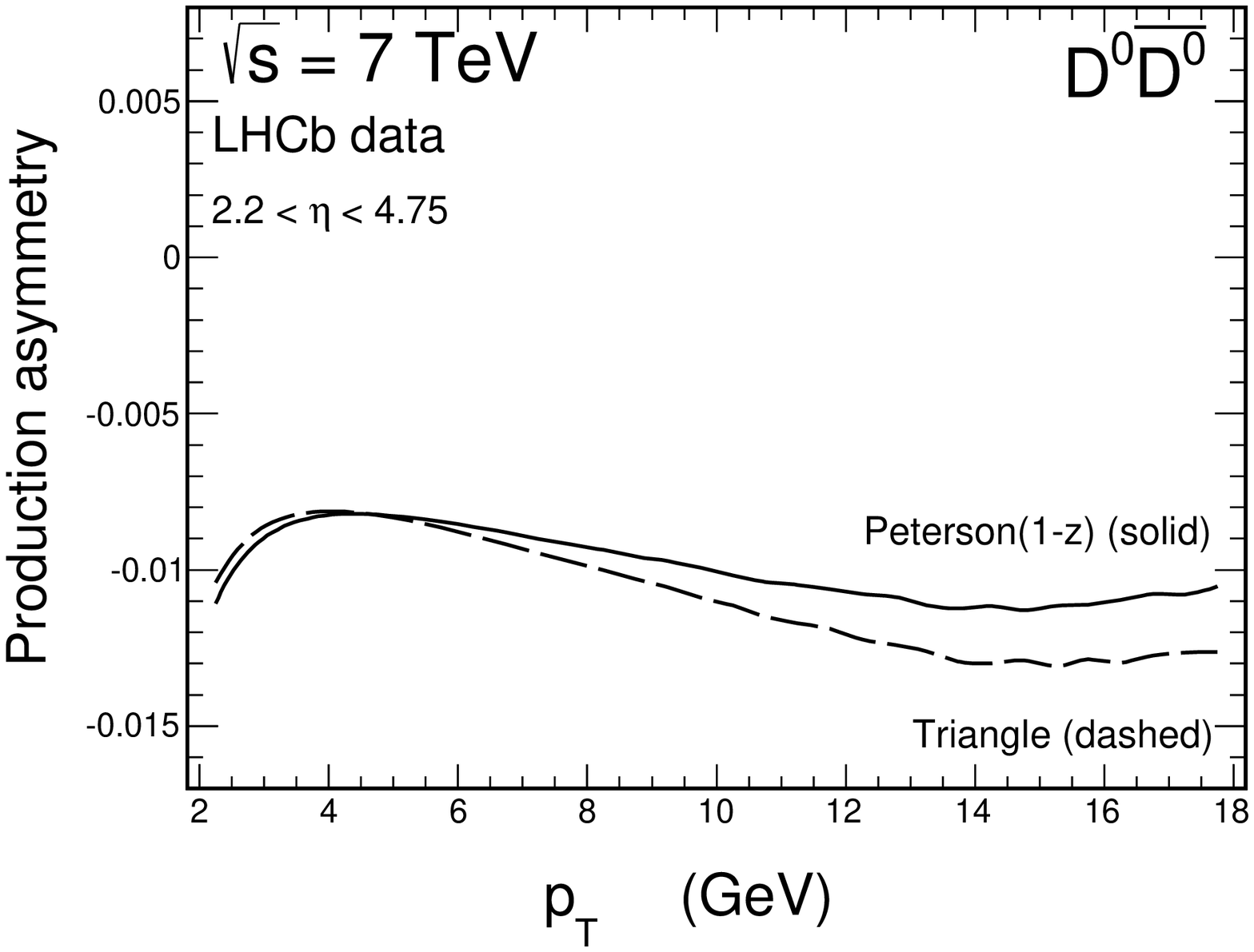}}
\end{minipage}
\begin{minipage}{0.47\textwidth}
  \centerline{\includegraphics[width=1.0\textwidth]{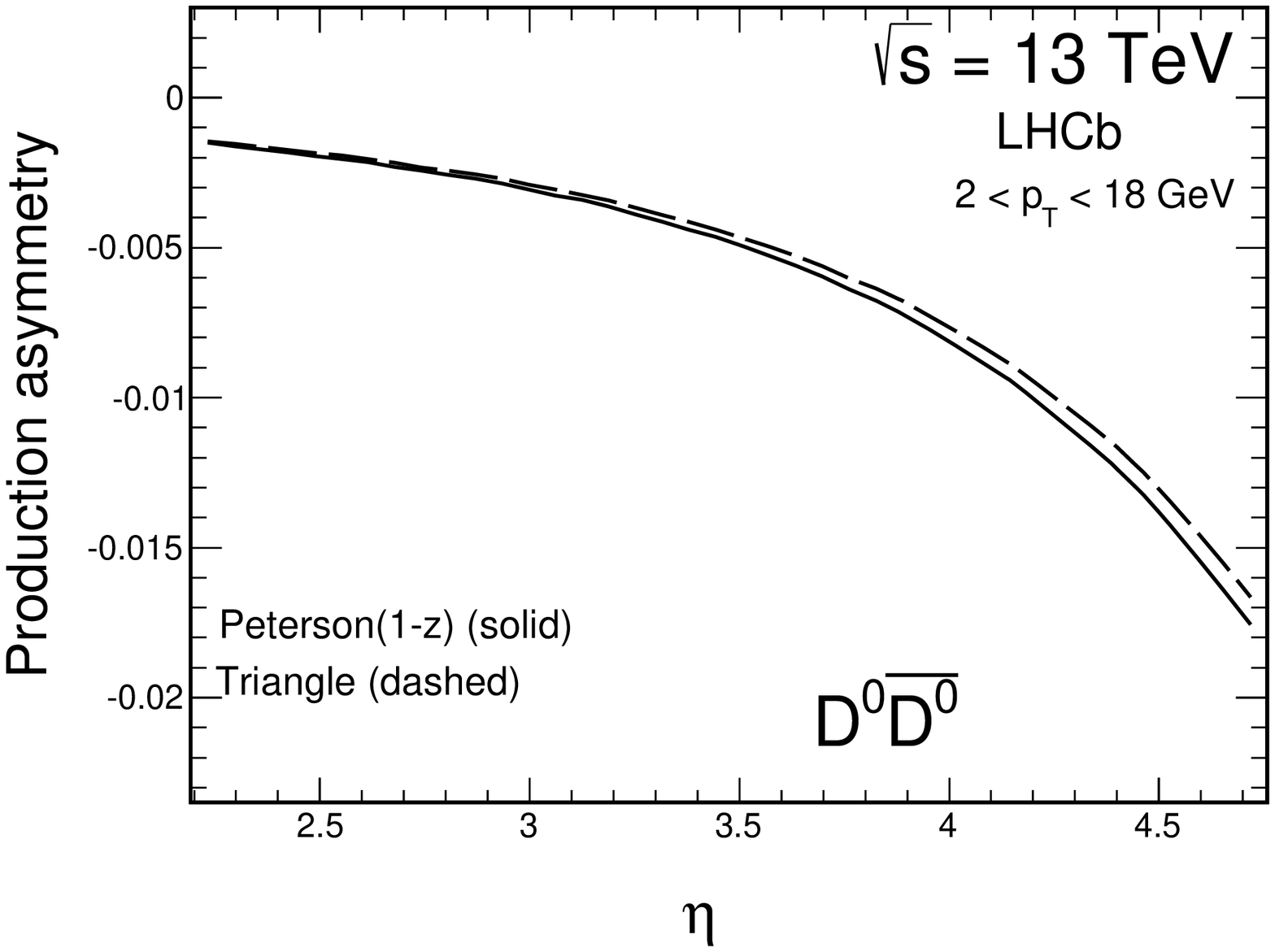}}
\end{minipage}
\hspace{0.5cm}
\begin{minipage}{0.47\textwidth}
  \centerline{\includegraphics[width=1.0\textwidth]{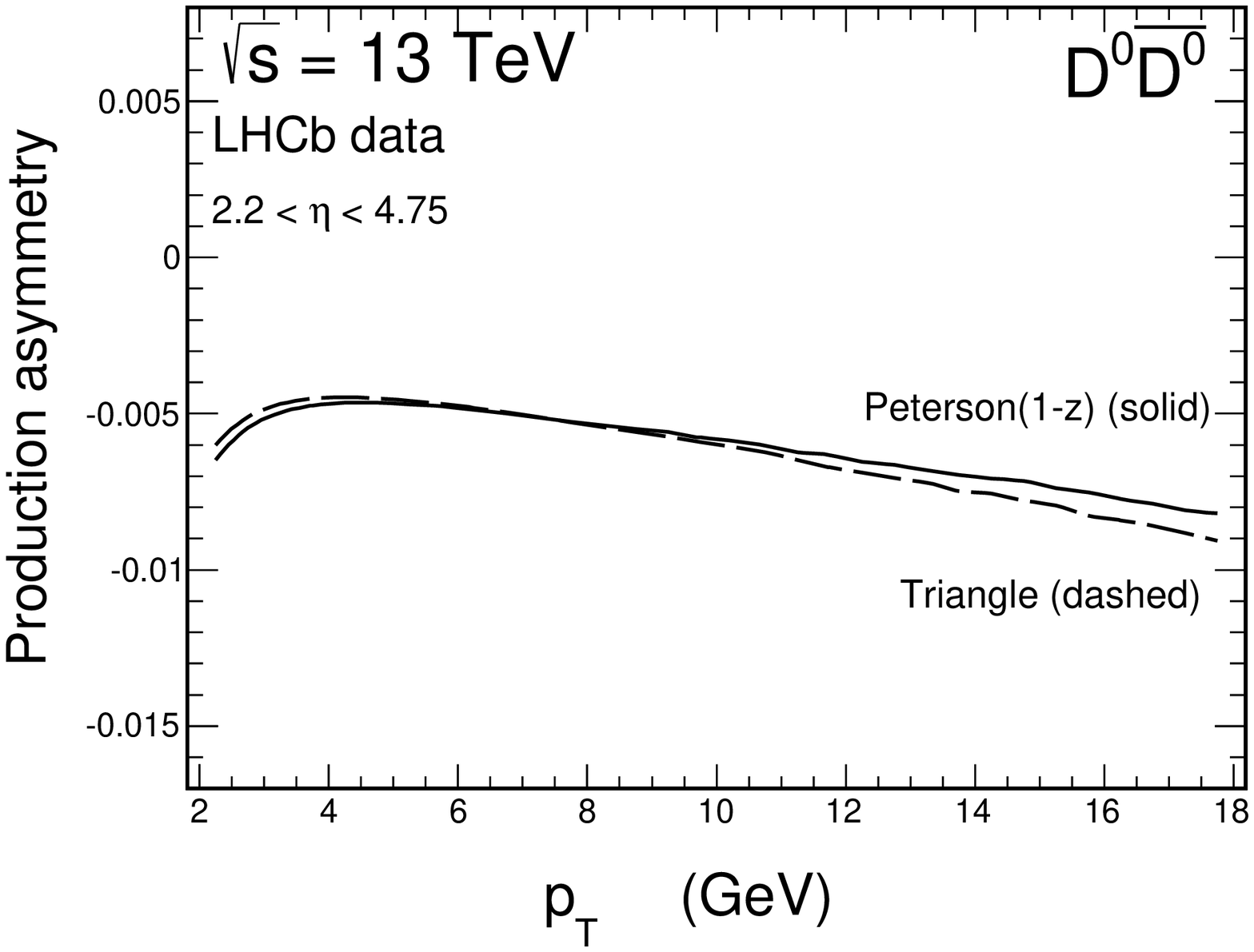}}
\end{minipage}
  \caption{
\small $A_{D^0/{\bar D}^0}$ production asymmetry relevant for
a possible LHCb collaboration measurement as
a function of $D$ meson pseudorapidity (left panel) and $D$ meson
transverse momentum (right panel).  The corresponding predictions for
$\sqrt{s}= 7$ and $13$ TeV 
are shown in the top and bottom panels, respectively.
}
\label{fig:LHCb_asymmetry_neutral}
\end{figure}

Having described the $A_{D^+/D^-}$ asymmetries for charged $D$ mesons we
wish to make predictions for $A_{D^0/{\bar D}^0}$ production asymmetries for neutral $D$ mesons.
According to our knowledge such asymmetries were never officially 
presented. The situation here is a bit more complicated due to $D^0$-${\bar D}^0$ mixing and resulting oscillations. 
Here we calculate production asymmetry.
In principle, the asymmetry may be (is) time dependent.
However, the oscillation time seems much longer than the life time of
$D^0/{\bar D}^0$ mesons, so it seems that the asymmetry could be
measured experimentally \textit{e.g.} by the LHCb collaboration.
This is very different for $B^0/{\bar B}^0$ mesons where the oscillation
time is rather short.
In Fig.~\ref{fig:LHCb_asymmetry_neutral} we show our predictions for
asymmetries for neutral $D$ mesons.
Slightly larger asymmetries are expected for $D^0/{\bar D}^0$ than
for charged $D^{\pm}$ mesons. $D^0/{\bar D}^0$ production symmetry is assumed in the LHCb studies of CP violation \cite{LHCb_A_CP}. Can such initial asymmetries have an influence
on the extracted $A_{CP}$ for neutral $D$ mesons?
This requires a separate dedicated study.

Now we shall make extrapolation to unmeasured regions.
Assuming flavour symmetry for direct production of pseudoscalar and
vector mesons (see Eq.~(\ref{flavour_symmerty_for_vector})) we shall make 
predictions also for $D^0$ and ${\bar D}^0$ production.

\subsection{\bm{$D \bar{D}$} asymmetry at lower energies}

\begin{figure}[!h]
\begin{minipage}{0.47\textwidth}
  \centerline{\includegraphics[width=1.0\textwidth]{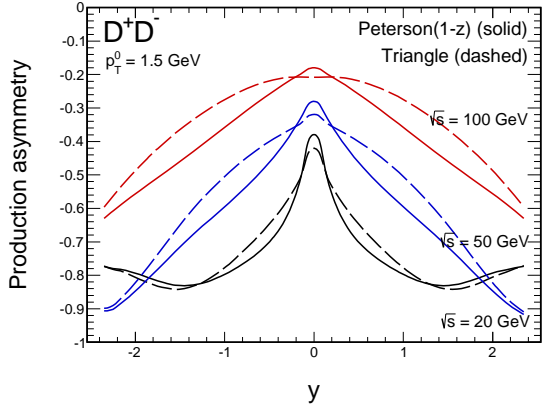}}
\end{minipage}
  \caption{
\small $A_{D^{+}D^{-}}(y)$ production asymmetry in proton-proton collisions for different $\sqrt{s}$
indicated in the figure.
}
\label{fig:assym_y_energy}
\end{figure}

The asymmetry in $D^+/D^-$ or $D^0/{\bar D}^0$ production is caused by
the relative amount of $q/{\bar q} \to D$ and $c/\bar c \to D$
fragmentation mechanisms.
Here we include all partonic processes with light quark/antiquark in the
final state.
In Fig.~\ref{fig:assym_y_energy} we show the asymmetries for three different
energies $\sqrt{s}$ = 20, 50, 100 GeV.
We observe that the asymmetry at the lower energies is much larger than that
for the LHC energies. Even at midrapidity $y \approx$ 0 we predict
sizeable asymmetries. Our rough predictions could be checked
experimentally at SPS \cite{SPS,NA61}, RHIC or at fixed target LHCb \cite{LHCb_fixedtarget}. Such experiments would allow to better pin down
the rather weakly constrained so far $q/{\bar q} \to D$ fragmentation 
functions.
Once this is done, a more realistic calculation for production of
prompt neutrinos in the atmosphere could be done.

The discussed by us mechanisms of subleading fragmentation of $D$ mesons
lead to enhanced production of $D$ mesons at lower energies.
In Table~\ref{tab:low_energies} we show as an example different
contributions to the production of $D^{+}/D^{-}$ mesons. The dominant at
high-energy $gg \to c \bar c$ mechanism gives only $13\%$ and $18\%$ 
for $\sqrt{s}=27$ and $39$ GeV, respectively and strongly underestimates the
NA27 \cite{AguilarBenitez:1988sb} and E743 \cite{Ammar:1988ta}
experimental data. Inclusion of the "subleading" contributions brings
theoretical calculations much closer to the experimental data.
We predict sizeable $D^{+}/D^{-}$ asymmetries at these low energies, see Fig.~\ref{fig:assym_y_energy}. 

\begin{table}[tb]%
\caption{Different contributions to the cross sections (in microbarns) 
for $D^{+}+D^{-}$ production at low energies. The results presented here
have been obtained with $p_{T}^{0} = 1.5$ GeV.}

\label{tab:low_energies}
\centering %
\begin{tabularx}{0.9\linewidth}{c c c}
\\[-4.ex] 
\toprule[0.1em] %
\\[-4.ex] 

\multirow{1}{4.5cm}{process:} & \multirow{1}{3.cm}{$\sqrt{s}=27$ GeV} & \multirow{1}{3.cm}{$\sqrt{s}=39$ GeV}\\ [+0.1ex]
\bottomrule[0.1em]
\multirow{1}{4.5cm}{$g^* g^* \to c\bar{c} \;\;\; (c/\bar{c} \to D^{\pm})$} & \multirow{1}{3.cm}{$1.52$} & \multirow{1}{3.cm}{$4.58$}\\ [+0.1ex]
\multirow{1}{4.5cm}{$q^* \bar q^* \to c\bar{c} \;\;\; (c/\bar{c} \to D^{\pm})$} & \multirow{1}{3.cm}{$0.08$} & \multirow{1}{3.cm}{$0.19$}\\ [+0.1ex]
\hline
\multirow{1}{4.5cm}{$g d \to g d \;\;\;\; (d \to D^{-})$} & \multirow{1}{3.cm}{$9.53$} & \multirow{1}{3.cm}{$13.89$}\\ [-0.2ex]
\multirow{1}{4.5cm}{$g \bar{d} \to g \bar{d} \;\;\;\; (\bar{d} \to D^{+})$} & \multirow{1}{3.cm}{$3.03$} & \multirow{1}{3.cm}{$4.78$}\\ [+0.1ex]
\hline
\multirow{1}{4.5cm}{$d d \to d d \;\;\;\; (d \to D^{-}) \times 2$} & \multirow{1}{3.cm}{$3.07$} & \multirow{1}{3.cm}{$4.29$}\\ [-0.2ex]
\multirow{1}{4.5cm}{$\bar{d} \bar{d} \to \bar{d} \bar{d} \;\;\;\; (\bar{d} \to D^{+}) \times 2$} & \multirow{1}{3.cm}{$0.29$} & \multirow{1}{3.cm}{$0.49$}\\ [-0.2ex]
\hline
\multirow{1}{4.5cm}{$\bar{d} d \to \bar{d} d  \;\;\;\; (d \to D^{-})$} & \multirow{1}{3.cm}{$0.58$} & \multirow{1}{3.cm}{$0.88$}\\ [-0.2ex]
\multirow{1}{4.5cm}{$d \bar{d} \to d \bar{d} \;\;\;\; (\bar{d} \to D^{+})$} & \multirow{1}{3.cm}{$0.58$} & \multirow{1}{3.cm}{$0.88$}\\ [-0.2ex]
\hline
\multirow{1}{4.5cm}{$u d \to u d \;\;\;\; (d \to D^{-})$} & \multirow{1}{3.cm}{$2.76$} & \multirow{1}{3.cm}{$3.72$}\\ [-0.2ex]
\multirow{1}{4.5cm}{$\bar{u} \bar{d} \to \bar{u} \bar{d} \;\;\;\; (\bar{d} \to D^{+})$} & \multirow{1}{3.cm}{$0.12$} & \multirow{1}{3.cm}{$0.19$}\\ [-0.2ex]
\hline
\multirow{1}{4.5cm}{$\bar{u} d \to \bar{u} d \;\;\;\; (d \to D^{-})$} & \multirow{1}{3.cm}{$0.40$} & \multirow{1}{3.cm}{$0.63$}\\ [-0.2ex]
\multirow{1}{4.5cm}{$u \bar{d} \to u \bar{d} \;\;\;\; (\bar{d} \to D^{+})$} & \multirow{1}{3.cm}{$0.97$} & \multirow{1}{3.cm}{$1.42$}\\ [-0.2ex]
\bottomrule[0.1em]
\multirow{1}{4.5cm}{theory predictions} & \multirow{1}{3.cm}{$22.93$} & \multirow{1}{3.cm}{$35.94$}\\ [-0.2ex]
\bottomrule[0.1em]
\multirow{1}{4.5cm}{experiment} & \multirow{1}{5.5cm}{NA27: $11.9 \pm 1.5$} & \multirow{1}{5.cm}{E743: $26 \pm 4 \pm 25\%$}\\ [-0.2ex]
\bottomrule[0.1em]
\end{tabularx}
\end{table}

The LHCb collaboration has an experience in measuring the asymmetry
in $D^+$ and $D^-$ production. It would be valueable to repeat such 
an analysis for fixed target experiment $p + ^{4}\!\mathrm{He}$ with gaseous target.
The data have been already collected.
The nuclear effects for $^{4}$He should not be too large.
Then the collision may be treated as a superposition of $p p$ and $p n$ 
collsions.
Neglecting the nuclear effects the differential cross section 
(in the collinear factorization approach) 
for production of $q/\bar q$ (particle 1) and 
associated parton (particle 2) can be written approximately as:
\begin{equation}
  \frac{d \sigma_{p\; ^4\! \mathrm{He}}}{d y_1 d y_2 d p_{T}} =
2 \frac{d \sigma_{pp}}{d y_1 d y_2 d p_{T}} +
2 \frac{d \sigma_{pn}}{d y_1 d y_2 d p_{T}} \; .
\label{nuclear_cross_section} 
\end{equation}
In the case of the second term we have to take into account
parton (quark/antiquark) distribution in neutron which can be obtained
from those in proton by assuming isospin symmetry between parton
distributions in the proton and neutron.
We are not interested in the distribution of gluons, that are treated
here as inactive in the production of $D$ mesons\footnote{A possible active role of gluons was discussed \textit{e.g.} in
Refs.~\cite{MSSS2016} in the context of double parton scattering.
Inclusion of the gluon fragmentation leads to much larger $\sigma_{\mathrm{eff}}$,
a parameter in the description of double-parton scattering.}.
Therefore an integration over gluon variables is performed as previously.

\begin{figure}[!h]
\begin{minipage}{0.47\textwidth}
  \centerline{\includegraphics[width=1.0\textwidth]{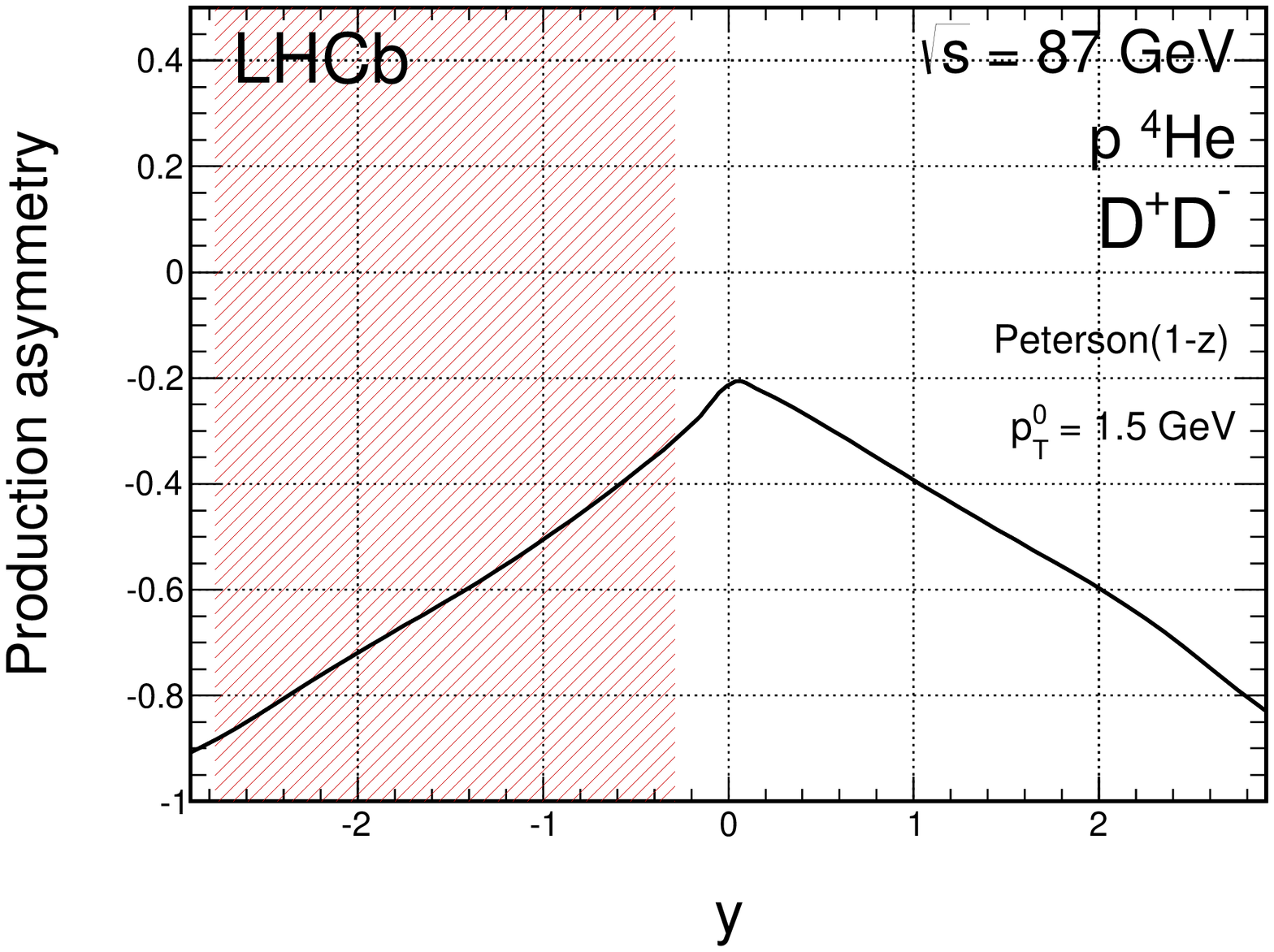}}
\end{minipage}
  \caption{
\small $A_{D^{+}D^{-}}(y)$ production asymmetry for the fixed target
$p+^4\!He$ reaction for $\sqrt{s}=87$ GeV.
}
\label{fig:assym_LHCb_fixed_target}
\end{figure}

In Fig.~\ref{fig:assym_LHCb_fixed_target} we present the relevant
predictions for the LHCb experiment.
Rather large asymmetries are predicted which could be addressed in the
forthcomming analysis of the fixed target experiment.

\subsection{Charge-to-neutral $D$ meson ratio}

\begin{figure}[!h]
\begin{minipage}{0.47\textwidth}
  \centerline{\includegraphics[width=1.0\textwidth]{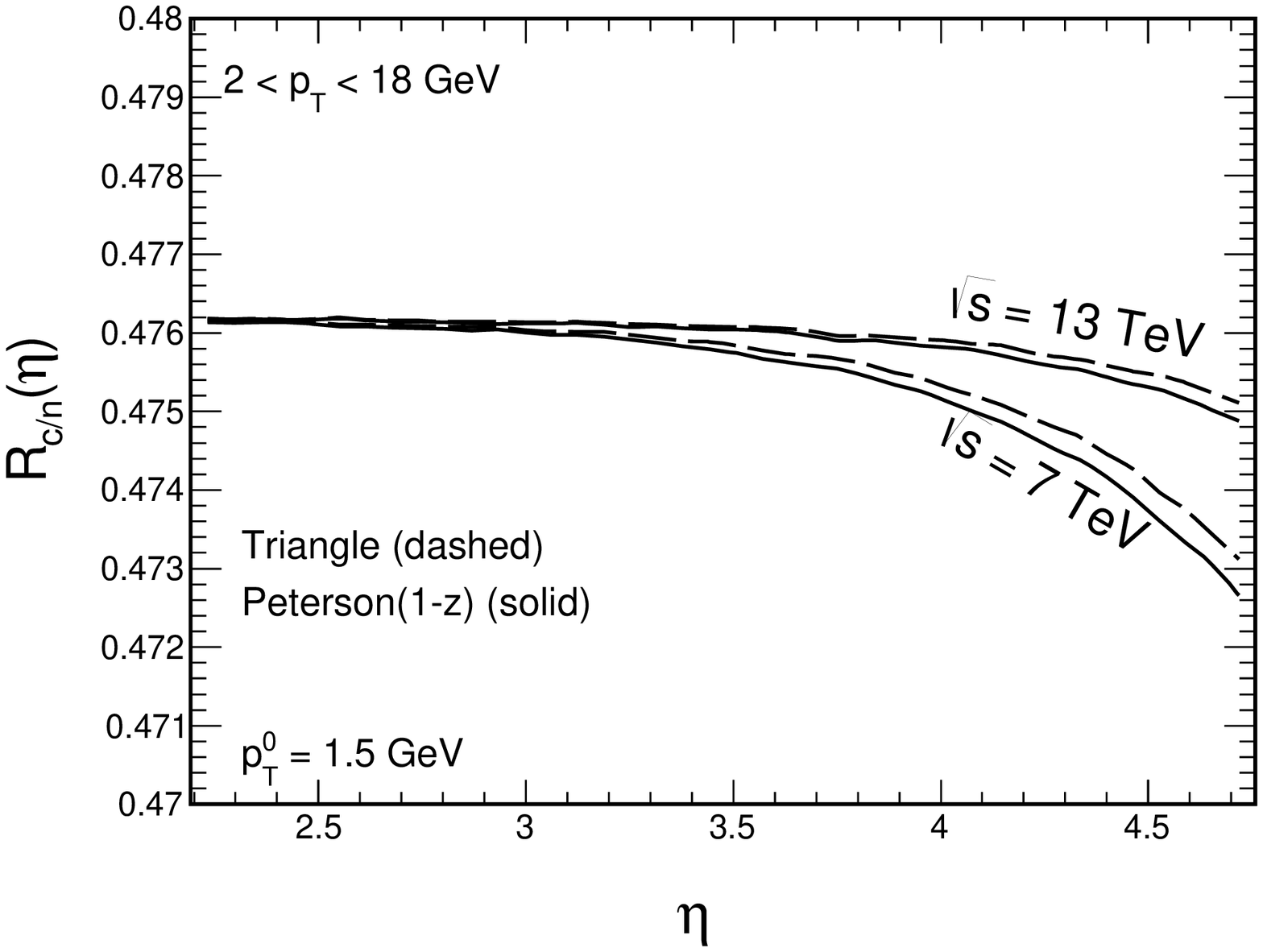}}
\end{minipage}
\begin{minipage}{0.47\textwidth}
  \centerline{\includegraphics[width=1.0\textwidth]{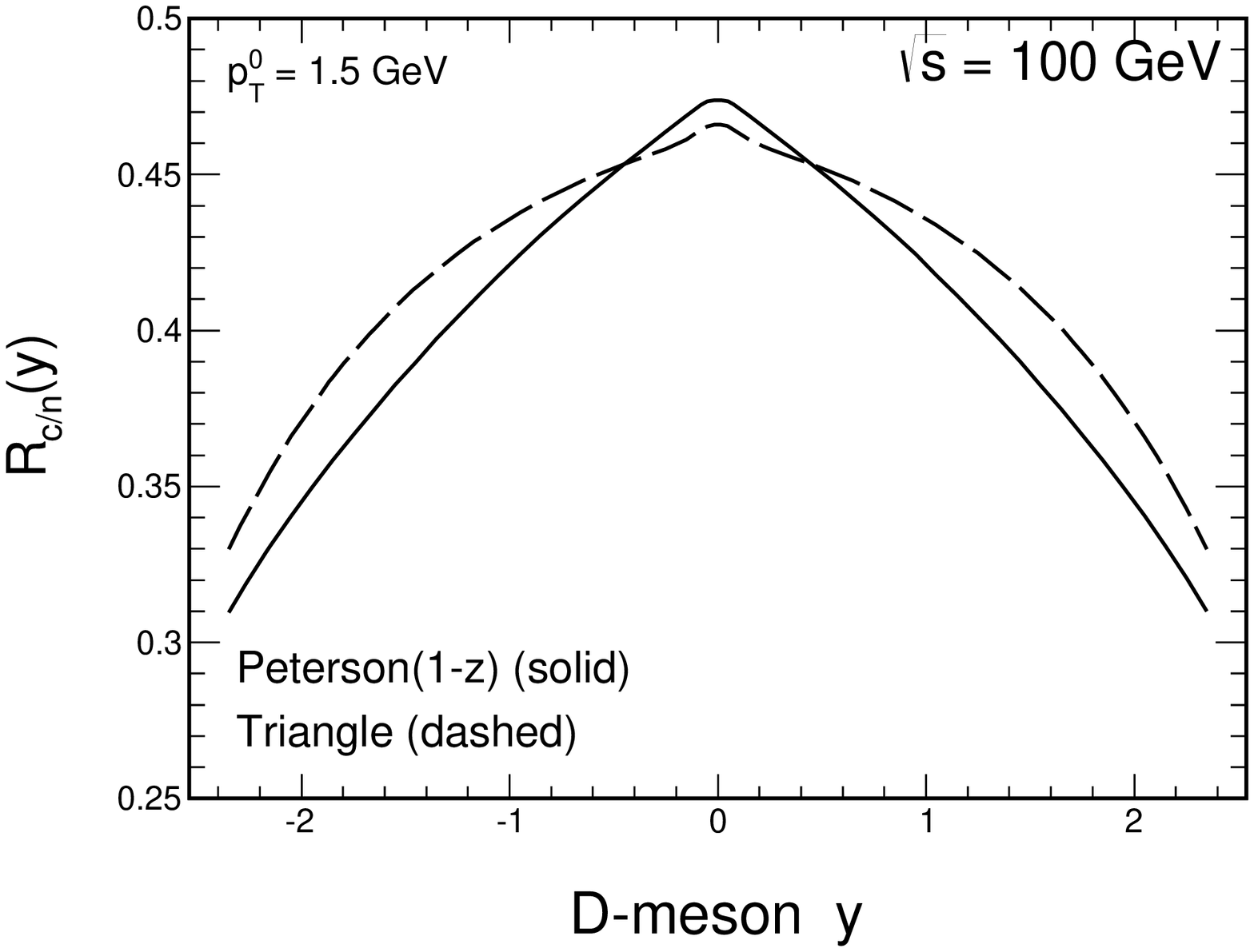}}
\end{minipage}
  \caption{
\small The $R_{c/n}$ ratio as a function of meson pseudorapidity for $\sqrt{s}= 7$ and $13$ TeV for the LHCb kinematics (left panel) and
as a function of meson rapidity for $\sqrt{s}$ = 100 GeV in the full phase-space (right panel). Only quark-gluon subleading components are included here.
}
\label{fig:R_cton}
\end{figure}

In the standard pQCD approach (production of $c/\bar c$ and
only $c/ \bar c \to D / \bar D$ fragmentation) the ratio defined as
\begin{equation}
R_{c/n} \equiv \frac{D^+ + D^-}{D^0 + {\bar D}^0} 
\label{R_cton}
\end{equation}
is a constant, independent of collision energy and rapidity (or $x_F$).
Inclusion of the subleading contribution changes the situation.
In Fig.~\ref{fig:R_cton} we show as an example the ratio as a function of 
meson pseudorapidity $\eta$ for LHC energies (left panel) and meson rapidity $y$ for $\sqrt{s}=100$ GeV (right panel), taking into account the
subleading contribution. At the LHC energies very small, difficult to measure, effect is found for the LHCb transverse momentum and pseudorapidity range. At $\sqrt{s}=100$ GeV we predict a strong rapidity dependence of the $R_{c/n}$ ratio. Perhaps fixed target experiments at the LHCb
could address the issue.

Identification of the dependence of $R_{c/n}$ on collision energy, 
rapidity or $x_F$ of $D$ mesons would be a good test of the considered 
here modeling and could better pin down the subleading fragmentation function.

\subsection{Resulting $D$ meson distributions
and possible consequenceses for prompt neutrino flux}

In this subsection we wish to show results relevant for high-energy
prompt atmospheric neutrinos. As discussed recently in Ref.~\cite{GMPS2017} 
a rather large $x_F \sim$ 0.5 region is important in this context.
The $d \sigma / d x_F$ distribution of mesons is the most appropriate distribution
in this context. For $x_F >$ 0.1 one can safely use the convolution 
formula from Eq.~(\ref{convolution}).

\begin{figure}[!h]
\begin{minipage}{0.47\textwidth}
  \centerline{\includegraphics[width=1.0\textwidth]{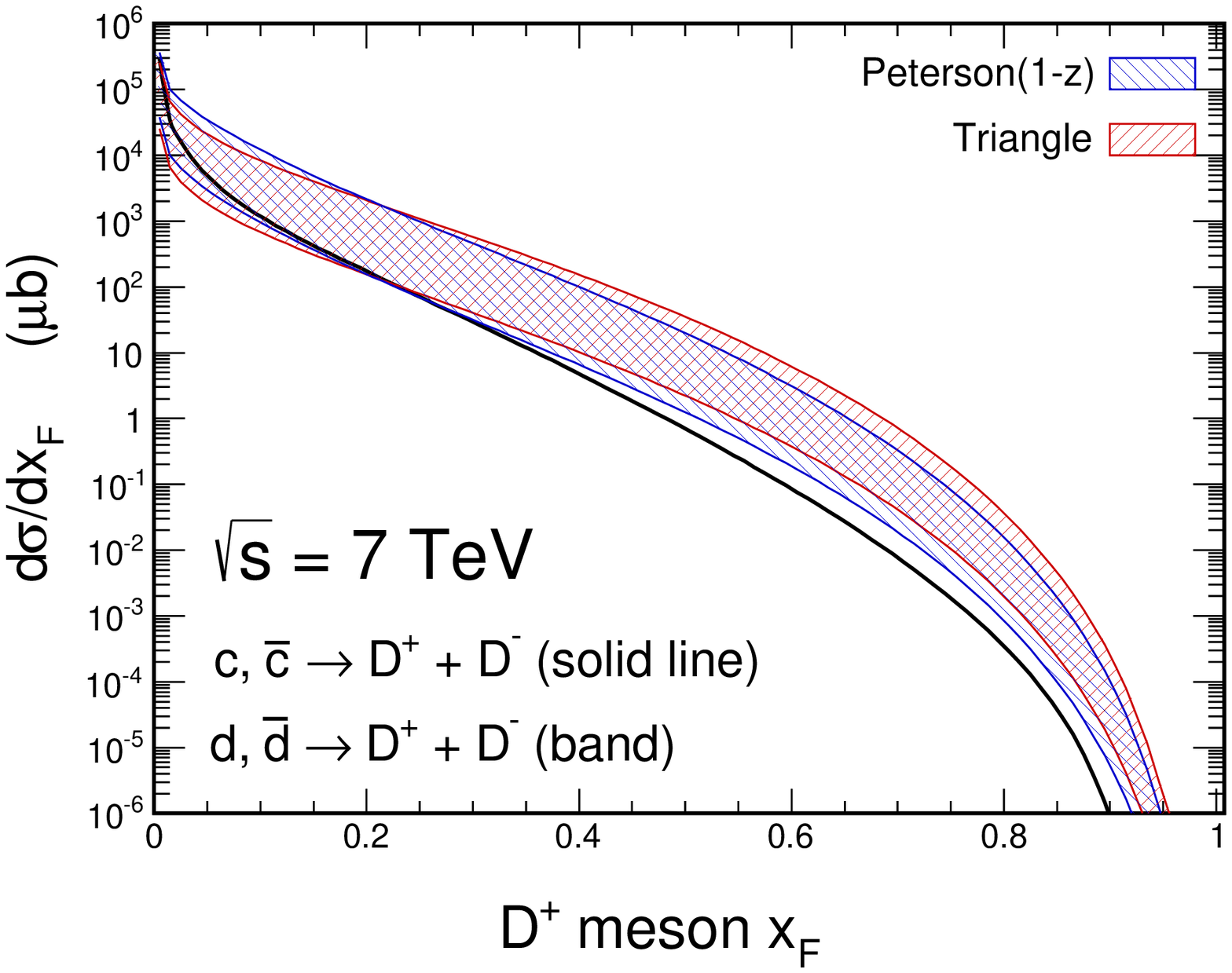}}
\end{minipage}
\begin{minipage}{0.47\textwidth}
  \centerline{\includegraphics[width=1.0\textwidth]{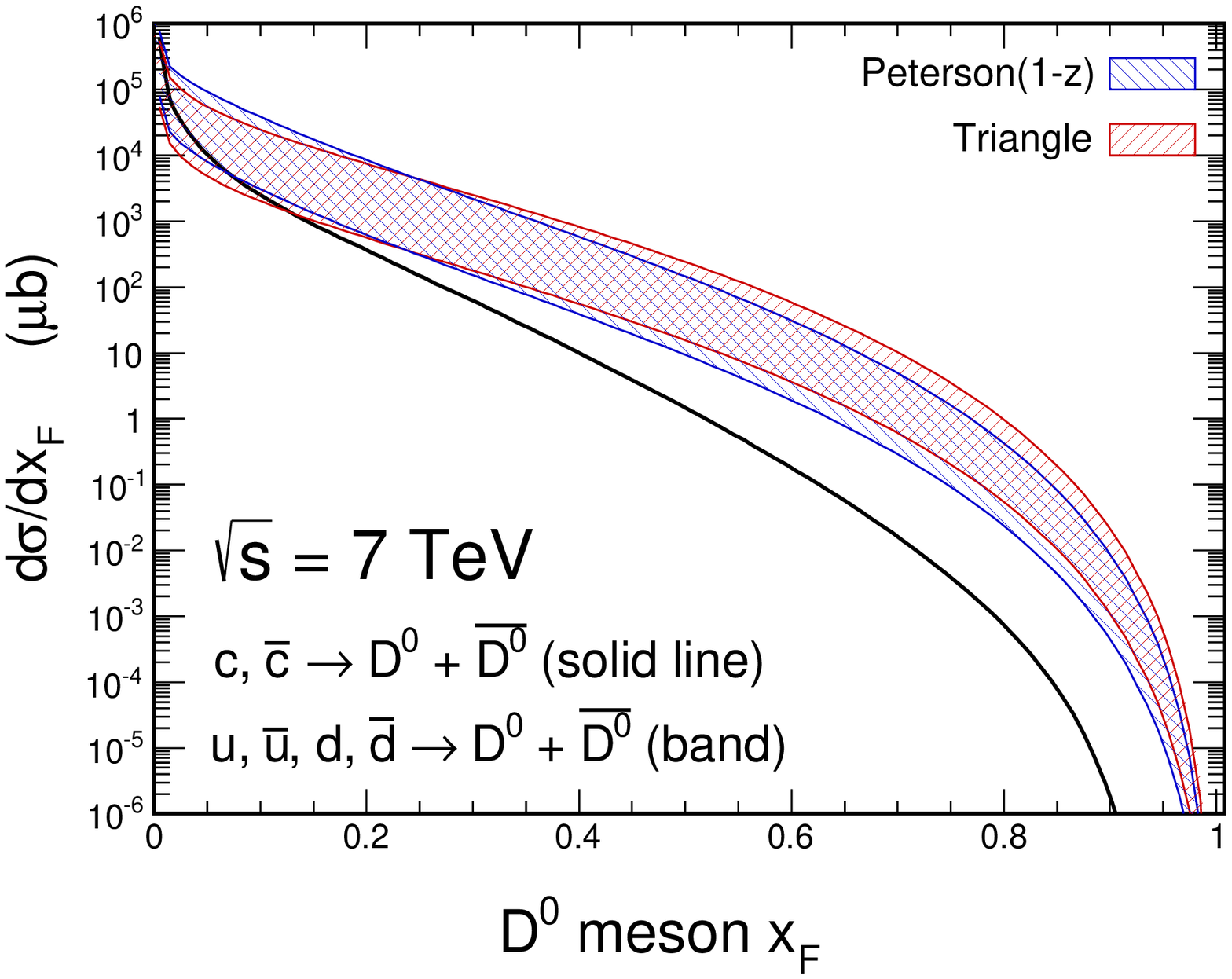}}
\end{minipage}
\begin{minipage}{0.47\textwidth}
  \centerline{\includegraphics[width=1.0\textwidth]{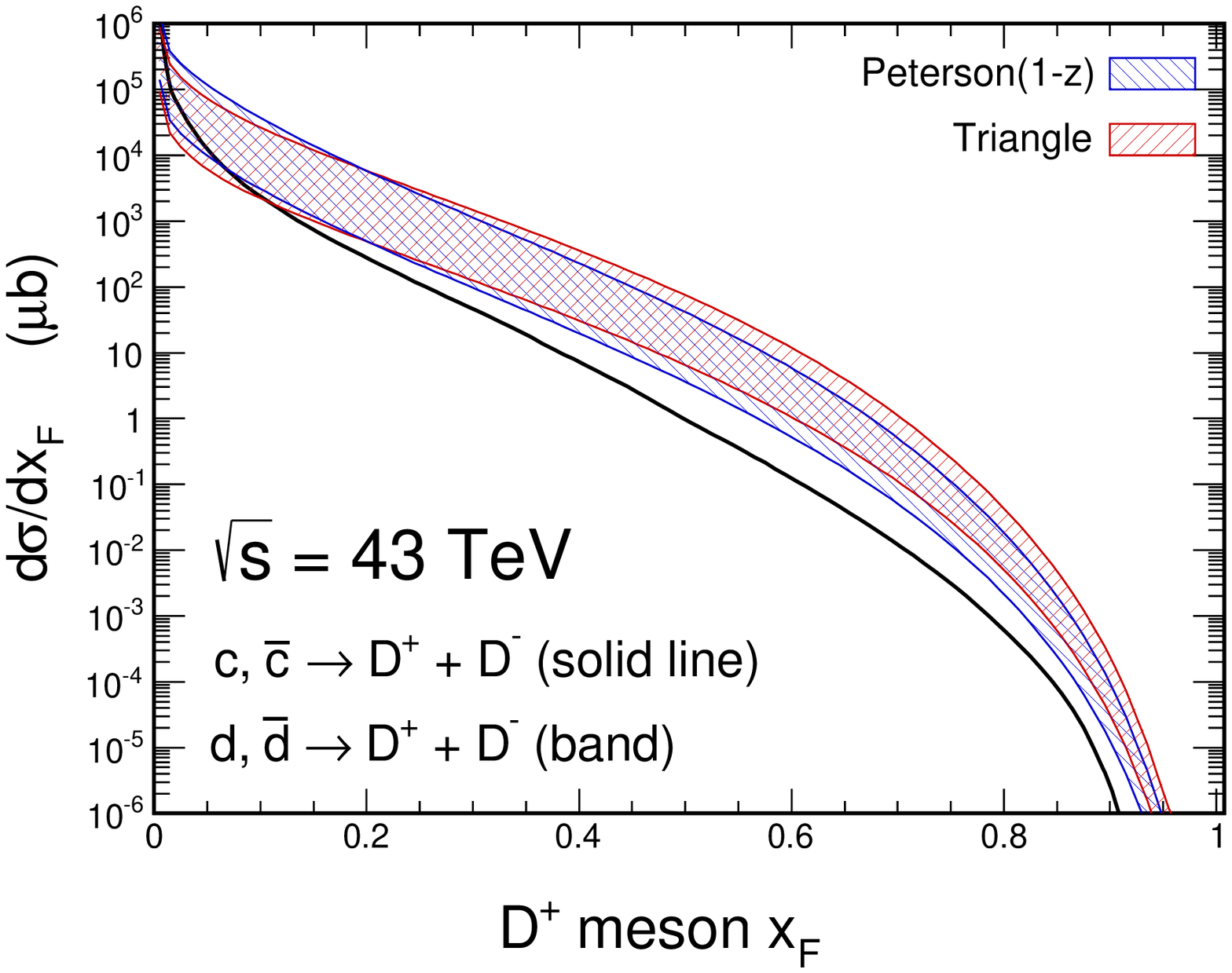}}
\end{minipage}
\begin{minipage}{0.47\textwidth}
  \centerline{\includegraphics[width=1.0\textwidth]{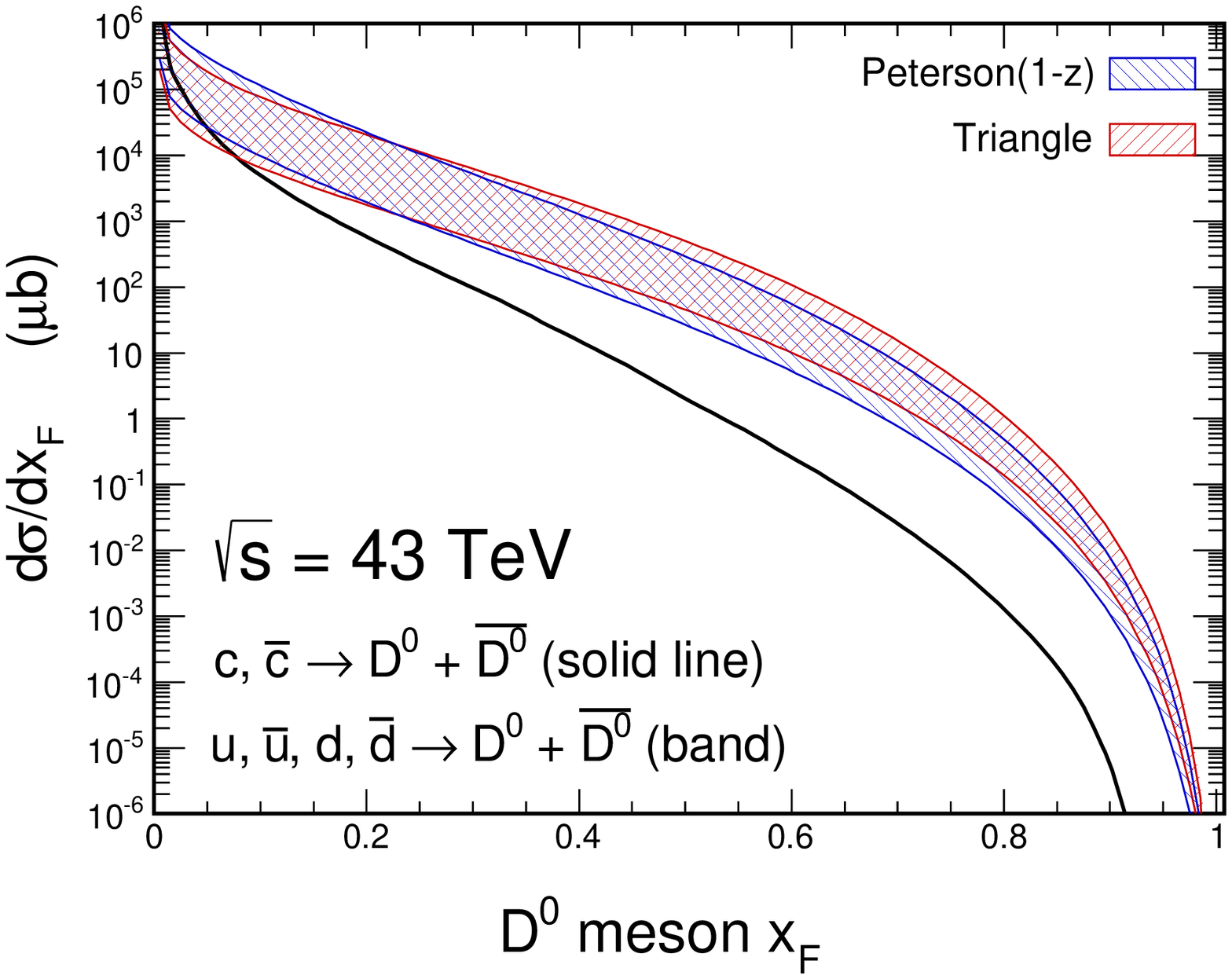}}
\end{minipage}
  \caption{
\small Distribution in $x_F$ for charged $D^+$+$D^-$ (left panel) and neutral 
$D^0$+${\bar D}^0$ (right panel) $D$ mesons from conventional 
(solid lines) and subleading (shaded bands) mechanisms.
The top panels are for $\sqrt{s}$ = 7 TeV and the bottom panels are for
$\sqrt{s}$ = 43 TeV.
}
\label{fig:dsig_dxf_D_mesons}
\end{figure}

\begin{figure}[!h]
\begin{minipage}{0.47\textwidth}
  \centerline{\includegraphics[width=1.0\textwidth]{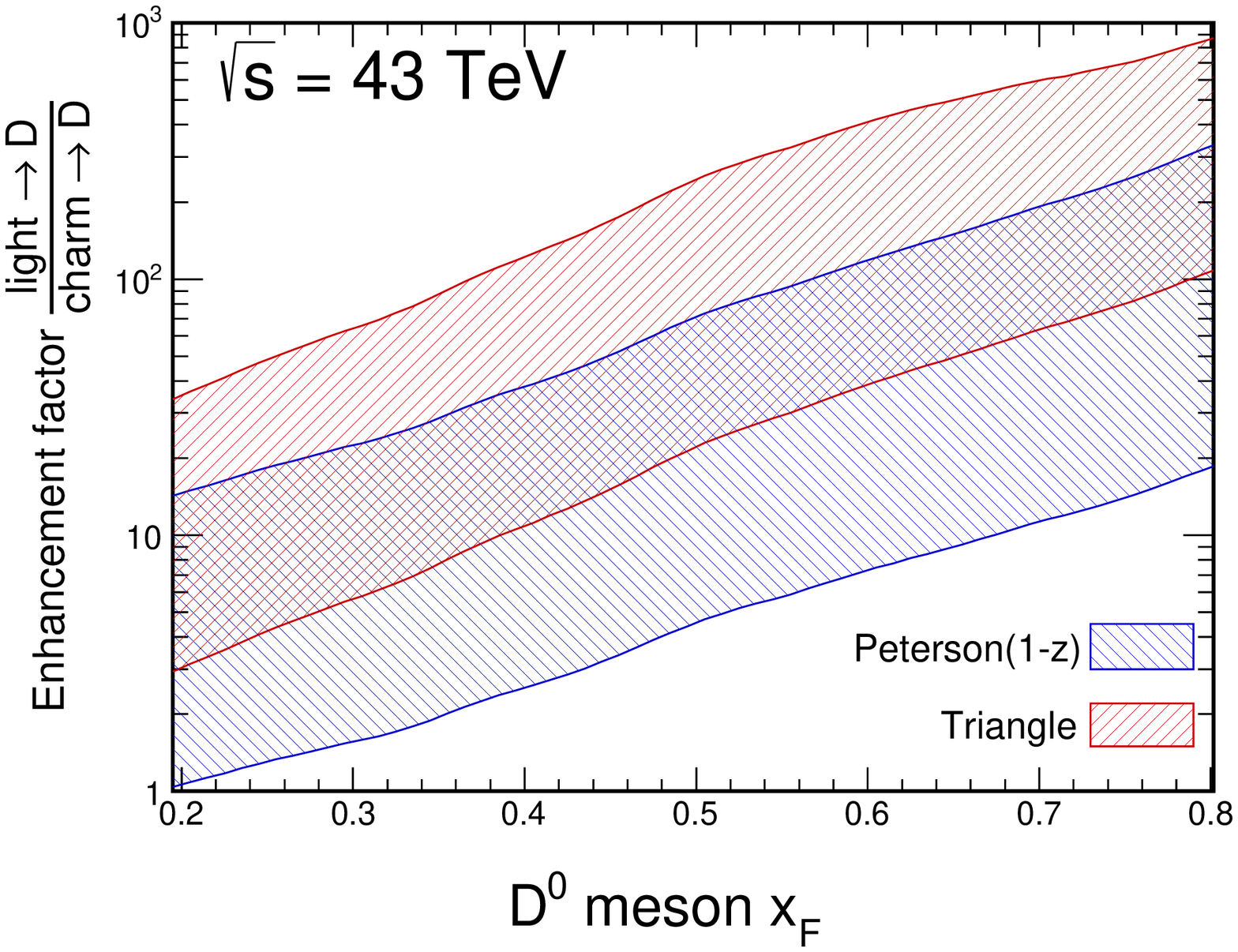}}
\end{minipage}
\begin{minipage}{0.47\textwidth}
  \centerline{\includegraphics[width=1.0\textwidth]{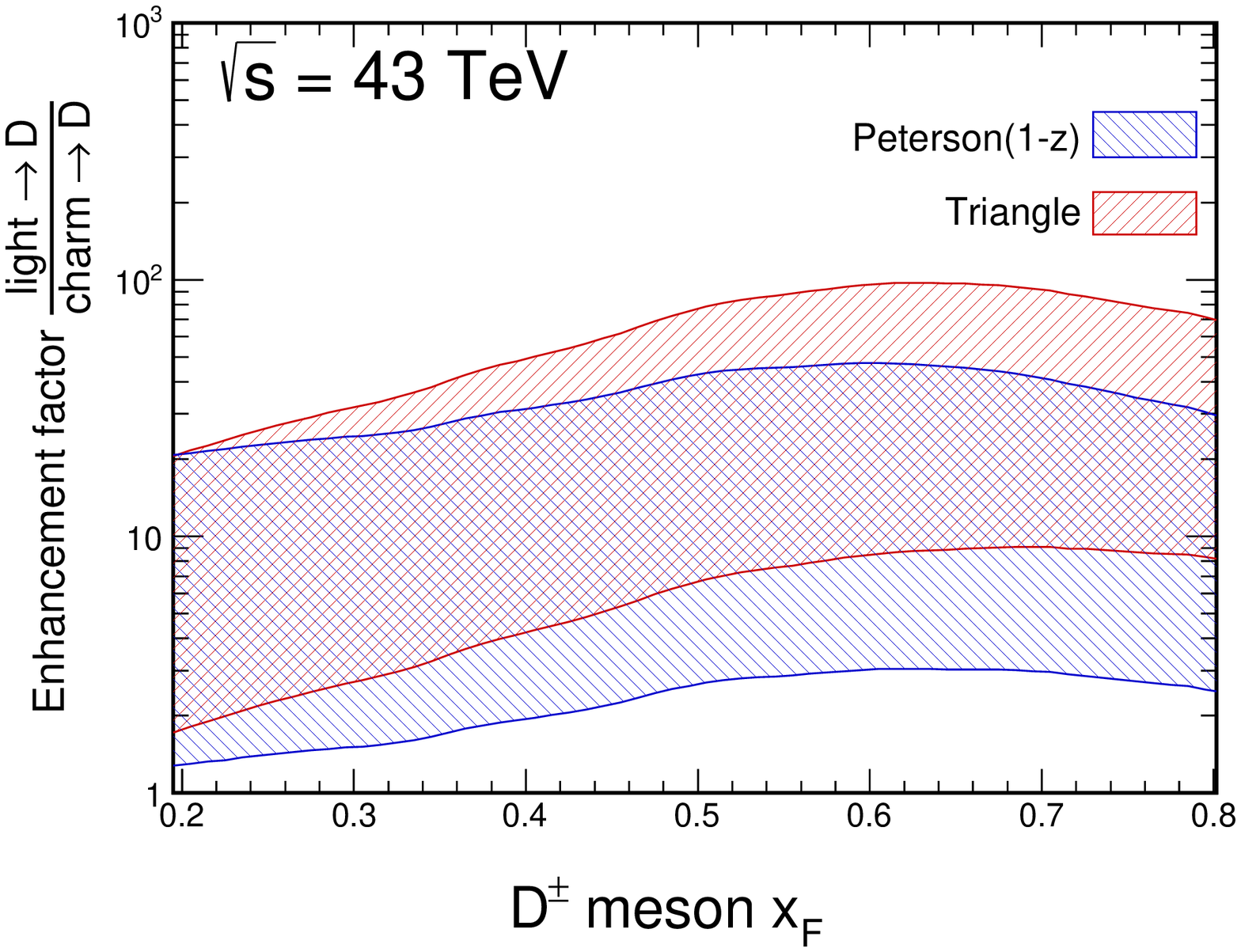}}
\end{minipage}
  \caption{
\small Enhancement factor for neutral (left panel) and charged (right panel) charm meson for $\sqrt{s}$ = 43 TeV.
}
\label{fig:enhancement}
\end{figure}

In Fig.~\ref{fig:dsig_dxf_D_mesons} we compare the two contributions:
(a) conventional one corresponding to $c \to D$ fragmentation and
(b) subleading one corresponding to $q \to D$ fragmentation,
for the sum of $D^{+}+D^{-}$ (left panels) and $D^{0}+\bar{D^{0}}$ (right panels) mesons.
While at small $x_F$ the conventional contribution dominates,
at large $x_F$ the situation reverses.
In addition we show the uncertainties bands where the upper and lower limits correspond to the
predictions for $p_{T}^{0}= 0.5$ and $1.5$ GeV, respectively.
The situation for both, $\sqrt{s} = 7$ (top panels) and $43$ TeV (bottom panels), energies is rather similar. The enhancement due
to the subleading contributions for neutral $D$ meson seems bigger than
that for charged $D$ mesons (see Fig.~\ref{fig:enhancement}). For example, for the triangle fragmentation functions, at $\sqrt{s}= 43$ TeV for $x_F \sim 0.5$ the cross
section for charged mesons ($D^+ + D^-$) is $3-15$ times bigger than 
for conventional approach while the cross section for neutral mesons
($D^0 + {\bar D}^0$) is $20-200$ times bigger.

\begin{figure}[!h]
\begin{minipage}{0.47\textwidth}
  \centerline{\includegraphics[width=1.0\textwidth]{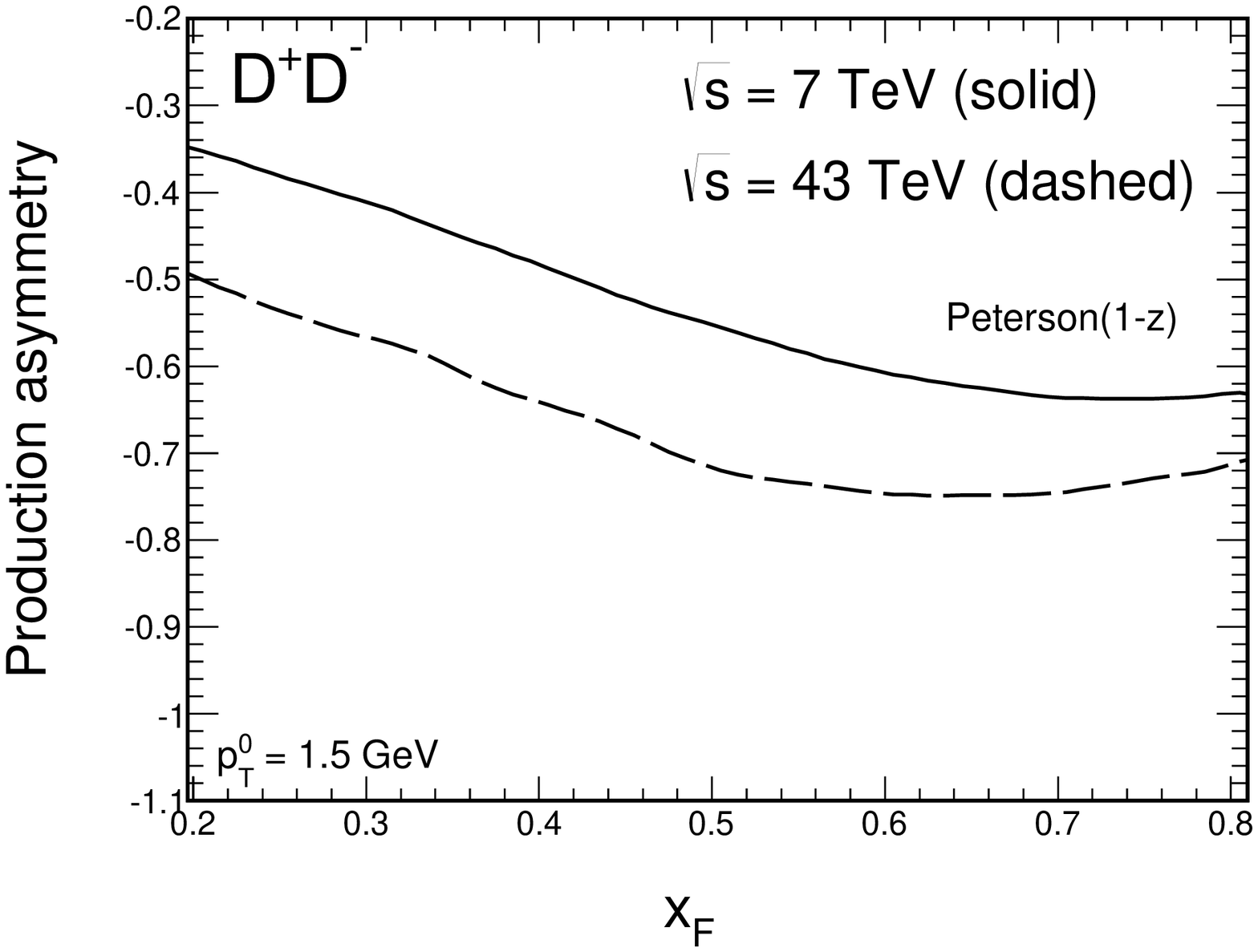}}
\end{minipage}
\begin{minipage}{0.47\textwidth}
  \centerline{\includegraphics[width=1.0\textwidth]{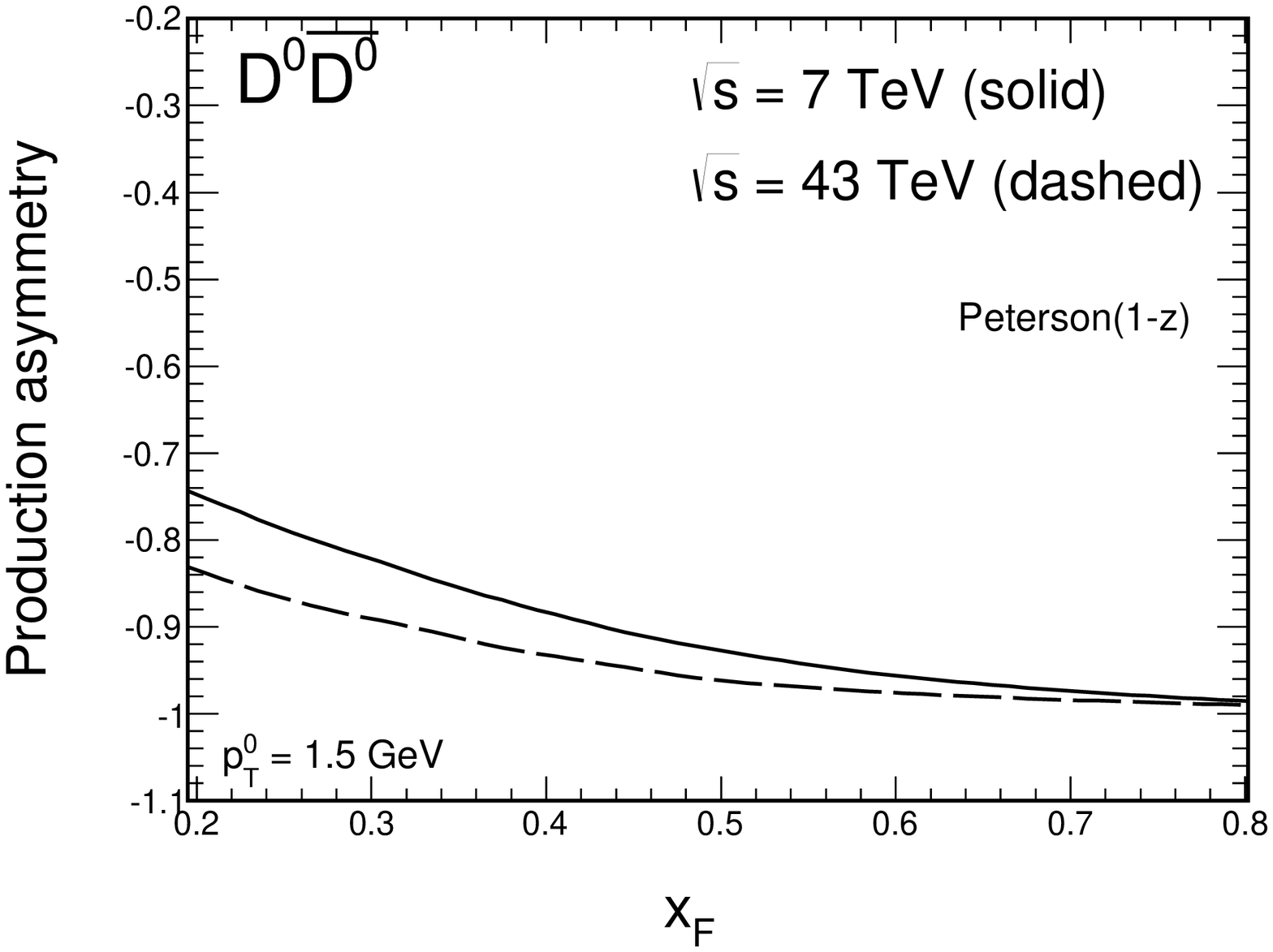}}
\end{minipage}
  \caption{
\small Production asymmetry as a function of $x_F$ for $D^+/D^-$ (left panel) and for $D^0/\bar{D^0}$ (right panel). The solid lines correspond to $\sqrt{s}=7$ TeV and the dashed lines correspond to $\sqrt{s}=43$ TeV.
The results are obtained with $p_{T}^{0} = 1.5$ GeV.
}
\label{fig:asymm_xF}
\end{figure}

We predict also asymmetry for $D^+/D^-$ and $D^0/{\bar D}^0$ production in the region of large $x_F$,
relevant for IceCube.
In Fig.~\ref{fig:asymm_xF} we show the asymmetry for the two large collision
energies. Within our model we predict larger asymmetries at larger energy in this kinematical domain.
Such asymmetries would lead to asymmetry in the production of neutrinos and antineutrinos.
We do not know whether this could or not be measured. 

The above results may have important consequences for large-energy 
atmospheric production which is not yet well understood background 
for cosmic (extraterrestial) neutrinos, claimed to be observed by 
the IceCube collaboration \cite{IceCube_flux}.
This will be a topic of a forthcomming analysis.
 

\section{Conclusions}

In the present paper we have discussed asymmetry in production
of $D^+$ and $D^-$ mesons in proton-proton collisions.
For a first time we have tried to understand whether the asymmetry 
observed by the LHCb collaboration can be understood within parton 
fragmentation picture, including light quark and antiquark fragmentation
functions.

The light quark/antiquark fragmentation to $D$ mesons arises naturally
within DGLAP evolution of fragmentation functions even assuming
vanishing fragmentation functions at some initial scale.
To understand the LHCb asymmetry we need, however,  
nonvanishing initial (for evolution) fragmentation functions.
Very small initial unfavoured fragmentation functions are sufficient
to describe the LHCb data. The details depend on functional form used.
The corresponding fragmentation probability for $q/{\bar q} \to D$ 
is very small, of the order of a fraction of 1\%, compared to 50 \%
for $c/{\bar c} \to D$ fragmentation.

Having described the asymmetry for charged $D$ mesons we have made
predictions for similar asymmetry for neutral $D$ mesons.
Nonzero asymmetries have been predicted.
This asymmetry may be, however, a bit more difficult to measure
due to $D^0-{\bar D}^0$ oscillations confirmed recently experimentally.

Furthermore we have predicted large contribution 
of the light quark/antiquark fragmentation to $D$ mesons at large 
$x_F$, which exeeds the conventional $c/{\bar c} \to D$ contribution. 


The predicted large contributions of $D$ mesons at large $x_F$ have 
important consequences for prompt neutrino flux at large neutrino energies, 
relevant for the IceCube measurements.
We have found that the contribution of the unfavoured fragmentation is
much more important than the conventional one for large
neutrino/antineutrino energies $E_{\nu} >$ 10$^{5}$ GeV. 

We have calculated in addition the asymmetries for much lower energies
($\sqrt{s}$ = 20 -- 100 GeV), relevant for possible measurements
in a near future.
Much larger asymmetries have been predicted, compared to those measured
by the LHCb collaboration \cite{LHCb_asymmetry}, 
even at $y \approx$ 0.
The asymmetries are associated with an increased production of charm in
the $q/{\bar q}$ initiated hadronization.
We have quantified this effect by discussing corresponding asymmetries
and rapidity distributions.
The corresponding measurements at fixed target LHCb, RHIC,  and at SPS (NA61-SHINE)
would allow to pin down the ``new'' mechanisms.
Especially the SPS experiment could/should observe an enhanced
production of $D$ mesons. Even a factor of 5 enhancement is not
excluded at present.

We have also predicted a dependence of the ratio of the charged-to-neutral $D$ meson
cross sections as a function of collision energy, meson rapidity or
$x_F$. We wish to remind in this context that different $K$ factors,
relative to pQCD calculations, were found long ago
for charged and neutral $D$ meson (see Ref.~\cite{BMDL1998}).

Systematic studies of $D/\bar D$ asymmetries or the specific ratios at low energies may be
therefore (paradoxically) important to understand the high-energy
prompt component of the atmospheric neutrino flux.

\vspace{1cm}

{\bf Acknowledgments}

This study was partially
supported by the Polish National Science Center grant
DEC-2014/15/B/ST2/02528 and by the Center for Innovation and
Transfer of Natural Sciences and Engineering Knowledge in Rzesz{\'o}w.
A.S. thanks Jolanta Brodzicka and Marcin Chrz{\c{a}szcz for a discussion
of the LHCb experiments on $D$ and $B$ meson production.
We also thank Victor Goncalves for a discussion on prompt atmospheric
flux. The exchange of information on fixed target LHCb experiments
with Michael Winn, Emilie Maurice and Frederic Fleuret is acknowledged.
We also thank Antoni Marcinek and Pawe{\l} Staszel for similar information
concerining planned NA61-SHINE experiment.



\end{document}